\documentclass[12pt]{iopart}

\usepackage{graphics}
\begin{document}

\title{Pythagorean Triplets, Integral Apollonians and The Hofstadter Butterfly}

\author{Indubala I Satija}

\address{Department of Physics and Astronomy, George Mason University, Fairfax, VA 22030}
\ead{isatija@gmu.edu}
\vspace{10pt}
\begin{indented}
\item[]October 12,  2018
\end{indented}

\begin{abstract}

Hierarchical sets such as
the Pythagorean  triplets ($\cal{PT}$)  and the integral Apollonian gaskets ($\cal{IAG}$)
are  iconic mathematical sets  made up of integers that resonate with a wide spectrum of inquisitive minds. Here we show that these abstract objects are
related with a quantum fractal made up of integers, known as the {\it Hofstadter Butterfly}. The ``butterfly fractal"  describes  a {\it physical system}  of electrons in a crystal in a magnetic field, representing  exotic states of matter known as {\it integer quantum Hall } states. Integers of the butterfly are the quanta of Hall conductivity that appear in  a highly convoluted  form in the integers of the $\cal{PT}$ and the $\cal{IAG}$. 
Scaling properties of these integers, as we zoom into the self-similar butterfly fractal  are given by  a class of quadratic irrationals that  lace the butterfly in a highly intricate and orderly pattern, some describing a 
 {\it mathematical kaleidoscope}.  The number theoretical  aspects are all concealed in Lorentz transformations along the light cone in abstract Minkowski space  where  subset of these  are related to the celebrated {\it Pell's equation}.

\end{abstract}

\section{Introduction}

Starting with Pythagoras  around $300$ BC,  Diophantus of Alexandria around $200$ AD  and  Pierre de Fermat  around  $1630$,   integers have been the darlings of many mathematicians.
 Hierarchical integer sets like the
Pythagorean tree (made up of Pythagorean triplets), integer Apollonian gaskets (consisting of an infinite number of mutually kissing (i.e., tangent) circles  of {\it integer} curvatures (inverse radii) that are nested inside each other), sets of integers in Fermat's last theorem,  and solutions of Pell's equation continue to engage mathematicians\cite{TN}. 

It is no secret that physicists are also in love with these whole numbers
particularly when they emerge as {\it quanta},  such as the quantum numbers of angular momentum -- a phenomenon rooted in rotational  symmetry or in topological quantum numbers in exotic states of matter.   The latter case describes  the mysterious and fascinating  phenomenon in condensed matter physics, known as the quantum Hall effect (QHE)\cite{TKNN}.  Rooted in topology that is linked to Berry phase\cite{Berry,BP}, the quantum  Hall conductivity  $\sigma_H$  is integer multiple of $\frac{e^2}{h}$ where $e$ is the electronic charge and $h$ is Planck's constant. In other words, Hall conductance is quantized in units of two fundamental
constants. Unlike abstract integers of mathematicians, these quantum numbers in physics are {\it real}, that is, they can be measured in laboratories with extremely
high precision (one part in a billion).

A simple model  of  a two-dimensional lattice in an intense magnetic field describes  all possible integer quantum Hall states of non-interacting fermions \cite{PToday}.  The  graph of solutions of the model summarizing all possible energies of electrons resembles a butterfly and is referred as the Hofstadter butterfly\cite{Hof} . It is a quantum fractal that is made up of integers. The quantum mechanics of two competing periods -- the lattice and the cyclotron radius --  lies at the heart of the butterfly graph. However,  number theory plays a central role in this drama happening on the stage of a two dimensional lattice in a magnetic field. There is a  deep, intricate  and almost magical relationship between  the butterfly fractal and the hierarchical sets of $\cal{PT}$  and $\cal{IAG}$. Some of the results relating butterfly to Apollonian were discussed in earlier\cite{EPJ, book}.  

Richard Feynman once remarked\cite{RF} that  ``I don't know why number theory does not find application in physics. We seem to need the mathematics of functions of continuous variables, complex numbers, and abstract algebra". As narrated here, the butterfly story is centered around the number theory.
Even today, the importance of number theory in physics is often
 unappreciated by the physics community. Therefore, we begin this paper with  a brief introduction to some aspects of the number theory that are relevant to the problem describing electrons in a crystal that is subjected to magnetic field.
Our  discussion starts with  the $\cal{PT}$, their  geometrical interpretation in terms of  the Minkowski space and its relation to Pell's equation and $\cal{IAG}$.  
We will then describe the butterfly graph and how number theory provides a mathematical framework  to describe its recursive structure. 

\section{Pythagorean Triplets}

\begin{figure}
\resizebox{1.0\columnwidth}{!}{%
  \includegraphics{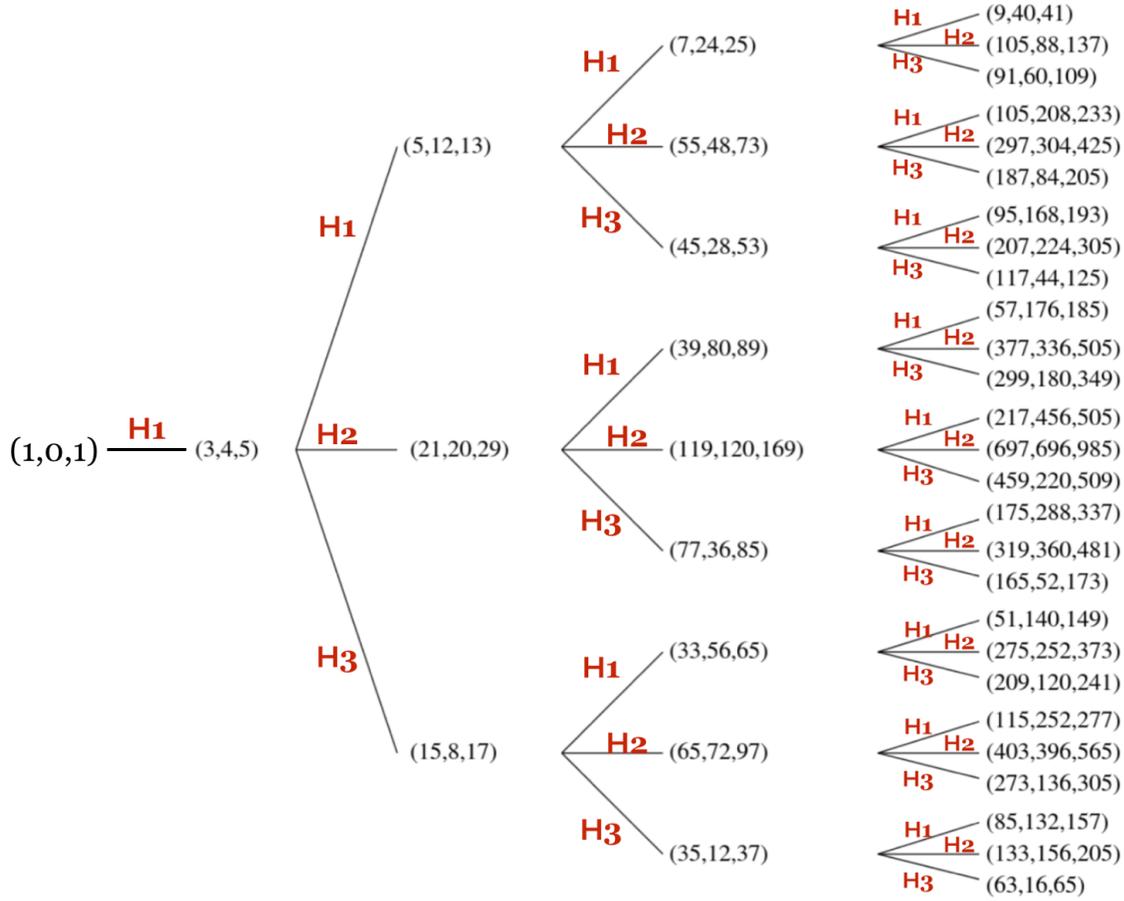} }
\leavevmode \caption{Four levels of the Pythagorean tree: Starting with $(3,4,5)$, all primitive $\cal{PT}$ can be generated by using three matrices $H_1,H_2,H_3$. As seen from the tree, the matrices $H_i$ preserve
the parity (even-oddness) of every member of a $\cal{PT}$.}
\label{PTnew}
\end{figure}

A Pythagorean triple is a set of three positive integers $(a, b, c)$   having the property that they can be respectively the two legs and the hypotenuse of a right triangle, thus satisfying the equation,

\begin{equation}
 a^2+b^2=c^2.
 \label{P}
 \end{equation}
 The triplet is said to be primitive if and only if a, b, and c share no common divisor.  

More than two thousand years ago, Euclid provided a recipe for generating such triplets of integers. 
Given a pair of integers $(m,n)$, $ m > n$,

\begin{equation}
a = m^2-n^2,\,\,\,\  b = 2mn, \,\,\,\ c = m^2+n^2
\label{Eu}
\end{equation}

In $1934$,  B. Berggren  discovered\cite{BPT}  that the set of all primitive Pythagorean triples has the structure of a rooted tree. In 1963,  this was rediscovered by the  Dutch  mathematician  F.J.M.  Barning   and  seven years later by  A. Hall\cite{Hall}  independently.  Using algebraic means, it was shown  that all primitive $\cal{PT}$ can be generated by three matrices  which we label as $H_1$, $H_2$ and $H_3$ as shown in Fig. (\ref{PTnew}). The three matrices are given by,

\begin{equation}
H_1=\left( \begin{array}{ccc} 1 & -2 & 2  \\    2 & -1 & 2  \\  2 & -2  & 3 \\ \end{array}\right),
H_2=\left( \begin{array}{ccc} 1 & 2 & 2  \\    2 & 1 & 2  \\  2 & 2  & 3 \\ \end{array}\right),
H_3=\left( \begin{array}{ccc} -1 & 2 & 2  \\    -2 & 1 & 2  \\  -2 & 2  & 3 \\ \end{array}\right) 
\label{H3}
\end{equation}

Using Euclid's recipe ( Eq. (\ref{Eu}) ), we can simplify this to three 
 $ 2\times 2$ matrices, acting on the pair $(m,n)$,  

\begin{equation}
h_1 =  \left( \begin{array}{cc} 1   & 2  \\  0 & 1   \\  \end{array}\right), \quad  h_2 =  \left( \begin{array}{cc} 2   & 1  \\  1 & 0   \\  \end{array}\right),\,\,\,\ h_3 =  \left( \begin{array}{cc} 2   & -1  \\  1 & 0   \\  \end{array}\right)
\label{h2}
\end{equation}

It is rather interesting to note that  Euclid's parametrization  of  the pair of integers $(m,n)$ as a $\cal{PT}$  is an example of a spinor\cite{JK}   because every  Pythagorean triplet   is a mapping,   $ z \rightarrow z^2$ for  any integer complex  number $ z$:
 
 \begin{eqnarray}
 z   & =  & m+ in\\
 z^2  &  =   &( m^2-n^2) +  i (2 m n)
   \equiv   a + i b, \, \, \ |z|^2 = c
 \label{pts}
\end{eqnarray}

Therefore, a rotation of $\pi$ of the parameter vector $z$  around the origin rotates $z^2$ by $2 \pi$ and this justifies the term ``Pythagorean spinor" for  the pair $(m,n)$.

 \subsection{  Pythagorean Triplets define a $2+1$ dimensional Minkowski space }
 
 There is a  geometric way to view the Pythagorean tree where the matrices ($H_1, H_2, H_3$) and ($h_1, h_2, h_3$) respectively correspond to the three and the two dimensional representation of the Lorentz group -- the group of transformations that  leave the quadratic form  $a^2+b^2-c^2=0$,  invariant\cite{JK}.
  
The space of triangles $(a,b,c)$ which we will from now onwards  denote as  $(n_x, n_y, n_t)$, can be interpreted  as a real 3-dimensional Minkowski space  (``space-time with the hypotenuse as the time" ) with a quadratic form

\begin{equation}
Q = n_x^2+n_y^2-n_t^2 =0.
\end{equation}

The right triangles are represented by ``light-like" (null) vectors. That is,  the Pythagorean triples are the integer null vectors in the light cone. The group of integer orthogonal matrices  (Lorentz transformations) permutes the set of Pythagorean triangles. 

The three $H$ matrices and their products are elements of the Lorentz group $O(2,1; Z)$. The $H$- matrices
 preserve the quadratic form $n_x^2+n_y^2-n_t^2=0$, that is $H_i^T G H = G $ with the pseudo-Euclidean matrix $G = diag(1,1,-1)$.
 Determinant of $H_1(h_1)$ and $H_3(h_3)$ are unity while that of $H_2(h_2)$ is equal to $-1$. 
It is easy to check that $H$ matrices 
can be split into a product of  matrices representing Lorentz boost  $L$ with velocity $v =\sqrt{\frac{8}{9}} $ , reflections $R_x$ and $R_y$ about the $x$ and $y$ axes and rotation $R_{\pi/4}$  in the $x-y$ plane as:

 \begin{equation} 
H_2 = R_y\,R_{\pi/4} L R_{\pi/4},\,\
H_3  =  R_y\, R_{\pi/4} LR_{\pi/4} R_x,\,\
 H_1= R_y\,R_{\pi/4}LR_{\pi/4} R_y 
 \end{equation}
  \subsection{Quadratic Irrationals as Eigenvalues of the $h$-matrices and Pell's Equation }

We note that the eigenvalues $E^{\pm}$ of $h_1$, $h_2$ and $h_3$ are respectively given by $(1,1)$, $(1+ \sqrt{2}, 1- \sqrt{2})$ and $(1,1)$. 
For any product of $h$ matrices,  the eigenvalues are irrationals of the form
 $2n \pm \sqrt{4n^2-1}$  or $2n \pm \sqrt{4n^2+1}$ where $ n=1, 2, 3.....$.  These two classes of irrationals are special class of  quadratic irrationals  as,
 
 \begin{eqnarray}
  2n \pm \sqrt{4n^2-1}  & = &  [2n+1; \overline{1, 2n} ]\\
  2n \pm \sqrt{4n^2+1} & =  & [2n;\overline{2n}],
  \label{cfrac}
  \end{eqnarray}
  
  where $[m, \overline{1, n}]$ represents periodic continued fraction with entries $1$ and $n$ and $m$ is the integer part of the irrational.  Two of the eigenvalues of the $H$ matrices are same  as the eigenvalues of the $h$ matrices.
  Their third eigenvalue is unity and signifies the existence of an invariant along any periodic path in the Pythagorean tree, relating it to Pell's  equation (more appropriately Pell-type equations)\cite{Pell}).
   
  Pell's  equation\cite{Pellbook}  is a quadratic Diophantine equation,
\begin{equation}
x^2 - s y^2 = d.
\label{pell}
\end{equation}
 For  a given pair of  integers $s$ and $d$, the integer pair $(x,y)$ is a solution of the equation. All  possible solutions  of the equation with $d=1$ correspond to rational approximants $\frac{x}{y}$ of the continued fraction expansion of the irrational $\sqrt{s}$ \cite{Pellbook}. Furthermore, all solutions with $d \ne 1 $ can be found from  the $d=1$ solutions once we obtain one solution for a given $d \ne 1 $.  
 
  The key to the relationship between the Pythagorean tree and the Pell's equation is the integer $d$ in Eq. (\ref{pell}) that corresponds to an invariant associated with a particular path in the tree characterized by a periodic string of $h$-matrices.
 For example, along the path $h_3 h_1$, the invariant is $d= 2n_y-n_t$. Expressing this in terms of Euclid parameters, we get $m^2-\sqrt{3}n^2 = d$, the Pell's equation with $x=m$ and $y=n$ and $s=3$.
The square root of the integer $s$ is the irrational number that appears in the eigenvalues of the string of $h$-matrices (denoted by $\hat{T})$. 
  Table (\ref{T2})  illustrates the correspondence between the Pythagorean tree and the Pell's equation with examples relating Euclid parameter $(m,n)$ with $(x,y)$, starting with the invariant $d$ expressed in terms of the triplets.
   
    \begin{table}
\begin{tabular}{| c | c |  c | c | c | c|  c|}
\hline
$\hat{T}$ \,\, &  $E_{\pm}$\,\,  &  $s$ \,\, & $d$ \,\   & $(x,y)$ \,\ & $\theta$   \\ \hline
$h_1$  \,\, & $   1 $\,\, &  $0$\,\  & $ n_y - n_t $ & $x = n$ \,\ &  $0.0$   \\ 
\hline
$h_3$  \,\, & $ 1 $\,\, & $0$\,\ &   $  n_x - n_t $ &  $x = m-n$ \,\ &  $\frac{\pi}{4}$   \\ 
\hline
$h_2$  \,\, & $ 1 \pm  \sqrt{2} $\,\, &  $2$\,\ &  $ n_x - n_y $ & $ x = m+n, y=m$\,\ &  $\frac{\pi}{8}$   \\ 
\hline
 $h_3 h_1 $  \,\, & $ 2 \pm \sqrt{3} $\,\, &  $3$ \,\ & $ 2n_y - n_t $  &  $x=m, y=n$ \,\ & $\frac{\pi}{6}$\\
\hline 
 $h_1 h_3 $  \,\, & $ 2 \pm \sqrt{3} $\,\, &  $3$ \,\ & $2n_x - n_t $  &  $x=m-2n, y= n$\,\ & $\frac{\pi}{12}$\\
 \hline
  $h_2 h_1 $  \,\, & $ 2 \pm \sqrt{5} $\,\, &  $5$ \,\  & $ 3n_y - 2n_t$  & $x=m, y=n$ \,\ &  $\frac{\pi}{7.47} $\\
   \hline
  $h_2 h_3 $  \,\, & $ 2 \pm \sqrt{5} $\,\, &  $5$ \,\ & $ 3n_x - 2n_t $  &  $x=2m-3n, y = n$ \,\ & $ \frac{\pi}{8.61}$\\
  \hline
  $h_1 h_2 $  \,\, & $ 2 \pm \sqrt{5} $\,\, &  $5$ \,\ & $2n_x - n_y$  &  $ x=n+2m, m$ \,\ &$ \frac{\pi}{13.55}$\\
   \hline
  $h_3 h_2 $  \,\, & $ 2 \pm \sqrt{5} $\,\, &  $5$ \,\ & $ n_x - 2n_y $  &  $x = m+2n, m$ \,\ &$ \frac{\pi}{5.67}$\\
   \hline
\end{tabular}
\caption{ Table lists examples showing correspondence between the Pythagorean tree described in terms of Euclid parameter $(m,n)$, the invariant $d$, the parameter $s$ and the solutions $(x,y)$ of the Pell's equation. The angle  $\theta$ specifies the eigenvector, $ ( \cos \theta, \sin \theta )$ of $\hat{T}$. 
  Note that every $T$ is associated with an invariant ( up to a sign ) $d$. }
     \label{T2}
\end{table}

\section{ Integral Apollonians }

\begin{figure}
\resizebox{1.0\columnwidth}{!}{%
  \includegraphics{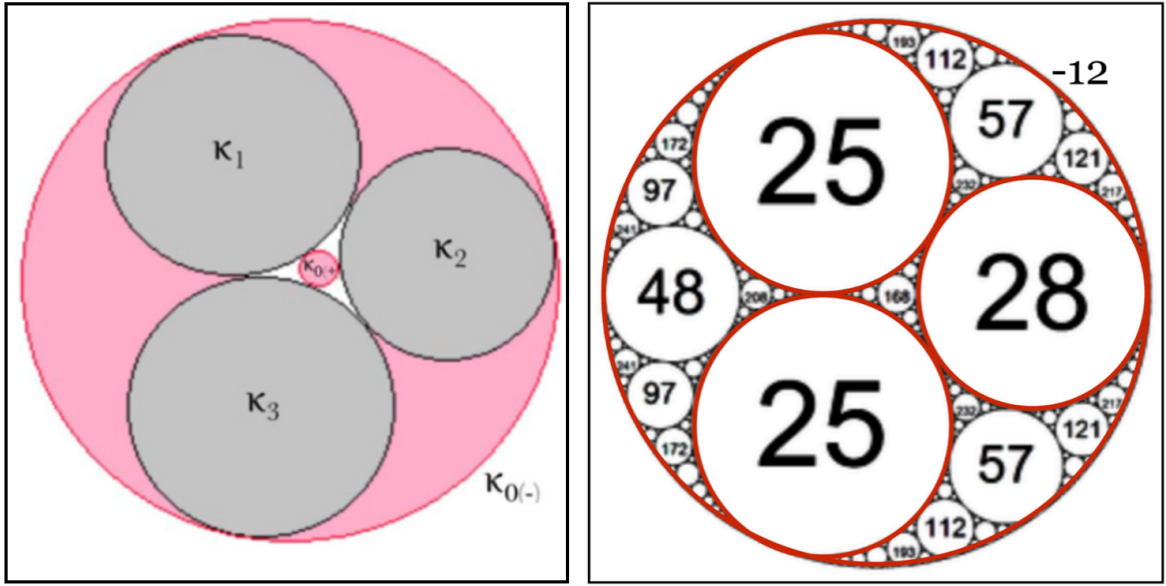} }
\leavevmode \caption{ Given three mutually kissing circles of curvatures $\kappa_1, \kappa_2, \kappa_3$, the left graph shows two solutions ( shown  in red)  $\kappa_0(+)$ and $\kappa_0(-)$ that are tangent to the three circles. 
The graph on the right shows an example of an $\cal{IAG}$ where all the circles have integer curvatures.}
\label{ap2}
\end{figure}

A configuration  of four mutually tangent or kissing circles as shown in  left panel in Fig. (\ref{ap2}) are known as Apollonian\cite{AP, book},
 named in honor of Apollonius of Perga born around 262 BC. 
A group of four numbers  $( \kappa_0, \kappa_1, \kappa_2, \kappa_3)$ that generate an Apollonian satisfy what is known as the Descartes theorem:

\begin{equation}
(\kappa_0^2+\kappa_1^2+\kappa_2^2+\kappa_3^2)= \frac{1}{2} (\kappa_0+\kappa_1+\kappa_2+\kappa_3)^2.
\label{DT}
\end{equation}

{\noindent The the four $\kappa_i$ are the {\it curvatures} --- that is to say, the reciprocals of the radii --- of the four circles.}\cite{RFD}. 

An immediate consequence of the Descartes' circle theorem is that given the curvatures of three  mutually kissing circles, there are two possible solutions for the 4th circle that is tangent to
the three circles. These two solutions, as shown in Fig. (\ref{ap2}),  correspond to the outermost and the innermost circle and are given by,
\begin{equation}
 \kappa_0 (\pm) = \kappa_1 + \kappa_2 + \kappa_3 \pm 2\sqrt{\kappa_1 \kappa_2 + \kappa_2 \kappa_3 + \kappa_3 \kappa_1}
 \label{DT1}
 \end{equation}
 
 The two solutions $\kappa_0(\pm)$ satisfy the linear equation,
$ \kappa_0(+) + \kappa_0(-) = 2 ( \kappa_1 + \kappa_2 + \kappa_3)$. This linear equation implies that
 if the outer and the three inner circles have integer curvatures, then all the inscribed circles 
in the curvilinear triangular regions between the circles,
have integer curvatures. Such configurations are known as $\cal{IAG}$  ( See right panel in Fig. (\ref{ap2})).

\begin{figure}
\resizebox{1.0\columnwidth}{!}{%
  \includegraphics{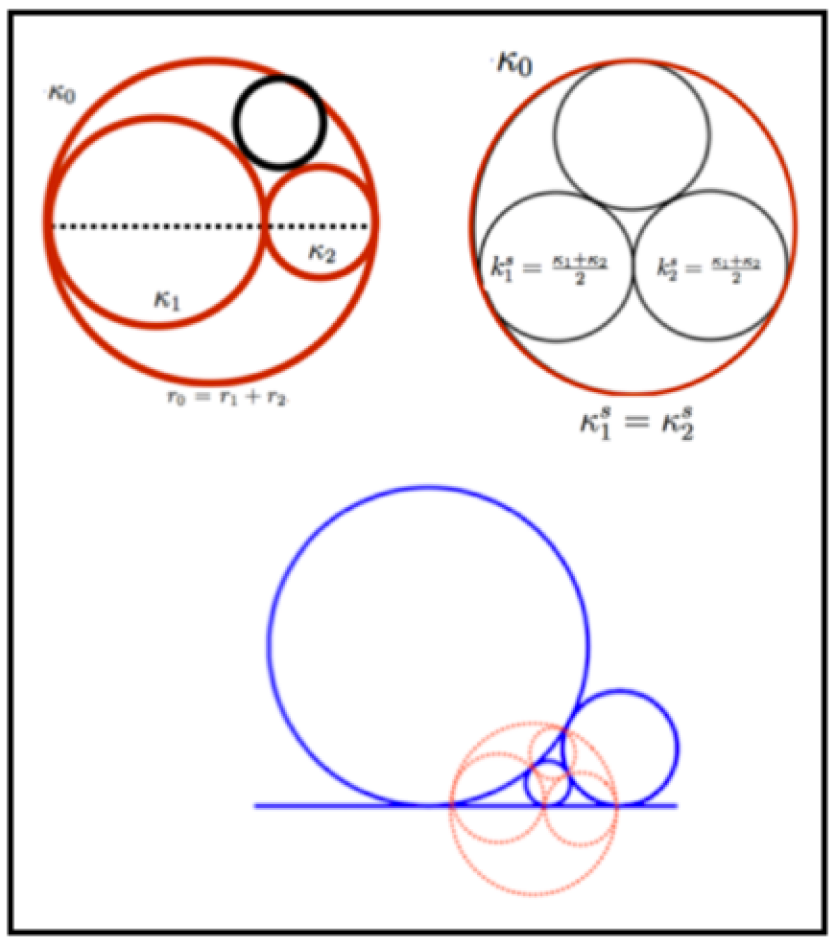} }
\leavevmode \caption{Figure shows three mathematically equivalent configurations: the collinear $\cal{IAG}$ and it's symmetrized partner (top row) are dual to the configuration of these kissing Ford circles -- the duality of four mutually kissing circles (shown in red) and their dual image  is (shown in blue). }
\label{ap}
\end{figure}

We now consider a subclass of Apollonians
 where  the centers of the three  of the four circles with curvatures $\kappa_0$,
$\kappa_1$ and $\kappa_2$ are collinear. In other words, their radii satisfy $r_0=r_1+r_2$ as shown in Fig. (\ref{ap}).  Interestingly, the integer curvatures of these Apollonians form a Pythagorean quadruplets, that is
$ \kappa_0^2+\kappa_1^2+\kappa_2^2=\kappa_3^2$.

 These Apollonians are related to two  other configurations as described  below in (I) and (II) and illustrated in Fig. (\ref{ap}).

(I) Collinear configurations of  Apollonian always have a partner -- a symmetrized configuration of four kissing circles
with curvatures $(\kappa^s_0, \kappa^s_1, \kappa_2^s, \kappa_3^s)$ where

\begin{eqnarray}
\left( \begin{array}{c}  \kappa_0^s \\  \kappa_1^s \\  \kappa_3^s \\ 0 \\ \end{array}\right) &   =  & \left( \begin{array}{cccc} 1 & 0 & 0 & 0 \\  0 & 1/2 & 1/2  &0 \\  -1 & 2 & 0 & 0 \\ -1 & 1 & 1 & -1 \\ \end{array}\right)   \left( \begin{array}{c}  \kappa_0 \\  \kappa_1 \\  \kappa_2 \\ \kappa_3 \\ \end{array}\right) 
\label{ks}
\end{eqnarray}

In other words,  $\kappa_0^s = \kappa_0$,  $ \kappa_1^s = \kappa_2^s = (\kappa_1+\kappa_2)/2$.  This is shown in the upper right panel of the Fig. (\ref{apt}).

(II)  A  collinear configurations of $\cal{IAG}$   is dual to  a mutually  kissing configuration of three circles that are tangent to $x$-axis, known as  the Ford circles\cite{Ford}. This ``duality"  transformation
 where the curvatures of the three kissing Ford circles are denoted as $(\kappa_L, \kappa_c, \kappa_R)$ is given by,

\begin{eqnarray}
\left(   \begin{array}{c}  -\kappa_0 \\  \kappa_1 \\  \kappa_2 \\   \kappa_3\end{array}\right)  =
   \frac{1}{2}  \left( \begin{array}{cccc } -1 & 1 & 1 & 1 \\    1 & -1   & 1 & 1  \\  1 & 1   & -1 & 1\\ 1 & 1 & 1 & -1 \\ \end{array}\right)   \left(  \begin{array}{c} \kappa_c \\  \kappa_R \\  \kappa_L  \\ 0   \\ \end{array}\right)
 \label{abc1}
\end{eqnarray}

The lower left panel of Fig. (\ref{apt})  illustrates the duality where  each circle in the dual set passes through three of the kissing points of the original set of circles, and the reverse holds as well.

\subsection{ Ford Circles and Farey Tree }

\begin{figure}
\resizebox{1.0\columnwidth}{!}{%
  \includegraphics{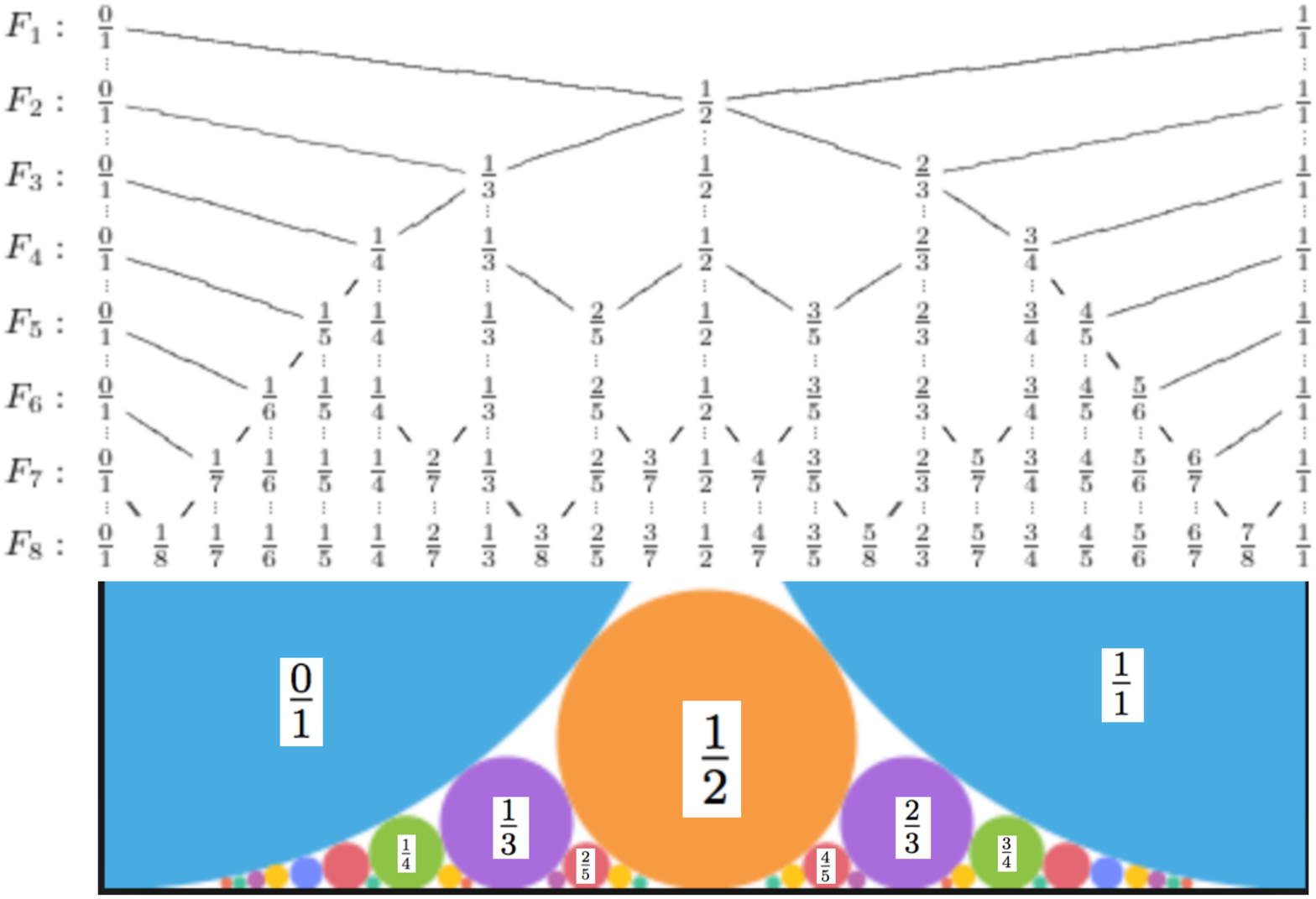} }
\leavevmode \caption{ Farey tree provides a systematic way of generating all primitive rationals.  Each successive row of the tree inherits all the rationals from the level above it, and is enriched with some new fractions (all of which lie between $0$ and $1$) made by combining certain neighbors in the preceding row using an operation called ``Farey addition". To combine two fractions $\frac{p_L}{q_L}$  and $\frac{p_R}{q_R}$ , one simply adds their numerators, and also their denominators, so that the Farey sum   is $\frac{p_L+p_R}{q_L+q_R}$. The bottom (colored part)  shows a visual representation of fractions as Ford circles.}
\label{FF}
\end{figure}

In his $1938$ paper\cite{Ford} entitled simply ``Fractions",  Ford proposed a geometrical visualization of fractions where
 every (primitive) fraction $\frac{p}{q}$ represented on the $x$-axis of the $xy$-plane can be associated with a circle -- Ford circle of radius $\frac{1}{2q^2}$ whose center is the point  $(x,y) = ( \frac{p}{q}, \frac{1}{2q^2})$.

Figure (\ref{FF}) shows the Ford circles along with their corresponding  (primitive) rational numbers, organized in a tree, known as a Farey tree.
Note that no two Ford circles ever intersect; the only way two Ford circles can meet is by {\it kissing} each other (being tangent at one point).  Two fractions whose Ford circles kiss are called
 {\it friendly} fractions. The condition for two fractions $\frac{p_L}{q_L}$ and $\frac{p_R}{q_R}$  to be friendly is given by
$| p_Lq_R-p_Rq_L | = 1$. 

\subsection{ Pythagorean Triplets meet Apollonians}

\begin{figure}
\resizebox{1.0\columnwidth}{!}{%
  \includegraphics{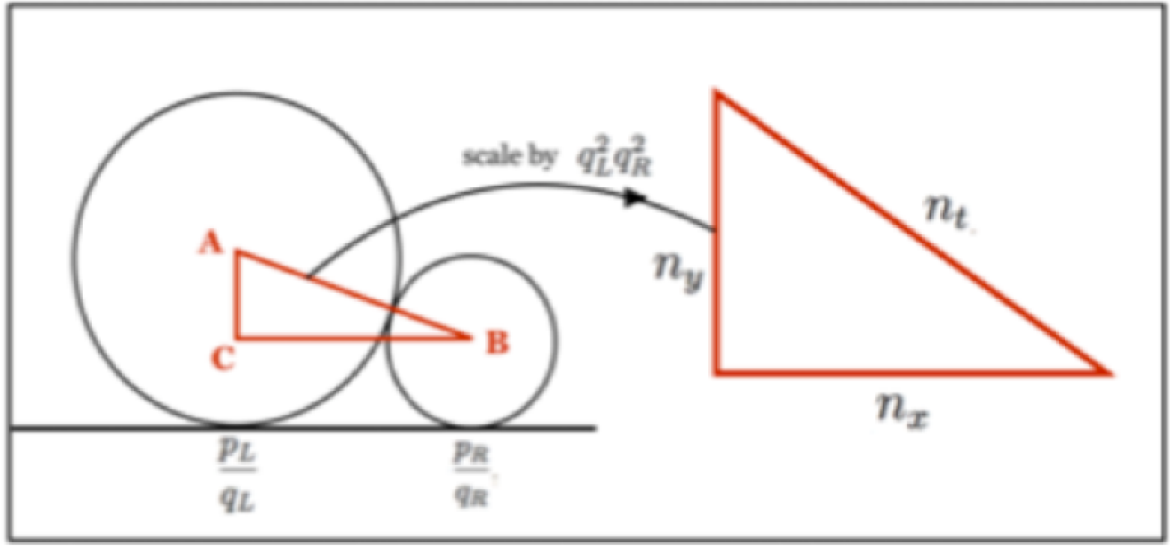} }
\leavevmode \caption{ Figure illustrates the mapping between the kissing Ford circles and
a Pythagorean triangle. After scaling, the sides of the triangle form a $\cal{PT}$ as described in the text.}
\label{apt}
\end{figure}

Given two kissing Ford circles,  we can draw a right angle triangle as shown in Fig. (\ref{apt}). For circles representing fractions $\frac{p_L}{q_L}$ and  $\frac{p_R}{q_R}$,
the coordinates of the centers of the circles are
 $(\frac{p_L}{q_L},  \frac{1}{2q_L^2})$ ,  $(\frac{p_R}{q_R},  \frac{1}{2q_R^2})$.
The three sides of the triangle are,
$( \frac{1}{q_L q_R},
 \frac{1}{2q_L^2} - \frac{1}{2q_L^2},
\frac{1}{2q_L^2} + \frac{1}{2q_R^2})$.
By scaling every side of this triangle by a factor $q_L^2 q_R^2$, the scaled right angle triangle will have sides $(q_L q_R,  \frac{q_R^2-q_L^2}{2},  \frac{q_R^2+q_L^2}{2})$ which are integers when $q_L$ and $q_R$ are odd and half-integers when $q_L,q_R$ have opposite parity.
 In the later case we  simply multiply by two to convert to a $\cal{PT}$.  Therefore,  for $q_c$ even ( $\kappa_0$ odd),  we have
 \begin{equation}
 n_x = \kappa_0,\,\ n_y= \frac{\kappa_2-\kappa_1}{2} ,\,\ n_t= \frac{\kappa_3-\kappa_0}{2}
 \end{equation}
 When $q_c$ is odd ( $\kappa_0$ even),
 \begin{equation}
 n_x = 2\kappa_0,\,\ n_y= \kappa_2-\kappa_1,\,\ n_t= \kappa_3-\kappa_0
 \end{equation}
 
 These Pythagorean triplets are related to the the curvatures of the three kissing Ford circles as,

\begin{eqnarray}
\left( \begin{array}{c}  \kappa_c \\  \kappa_R \\  \kappa_L \\ \end{array}\right) & = &  \left( \begin{array}{ccc} 1 & -1 & -1  \\    0 & 1 & -1  \\  0 & 1  & 1 \\ \end{array}\right)   \left( \begin{array}{c}  n_x \\  n_y \\  n_t  \\ \end{array}\right) 
\end{eqnarray}

\section{ The Hofstadter Butterfly}
 
 In 1974, few years before fractals became well-known, Douglas Hofstadter, then a physics graduate student at the University of Oregon, was trying to understand the quantum behavior of an electron in a crystal in the presence of an intense magnetic field. As he carried out his explorations by graphing the allowed energies of the electron as a function of the magnetic field\cite{Hof}, he discovered that the graph resembled a butterfly with a highly intricate recursive structure. It  consisted of nothing but copies of itself, nested infinitely deeply. Originally dubbed ``Gplot" or a ``picture of God", the graph is now known as the ``Hofstadter butterfly"\cite{DL}.
 
 \begin{figure}
\resizebox{1.0\columnwidth}{!}{%
  \includegraphics{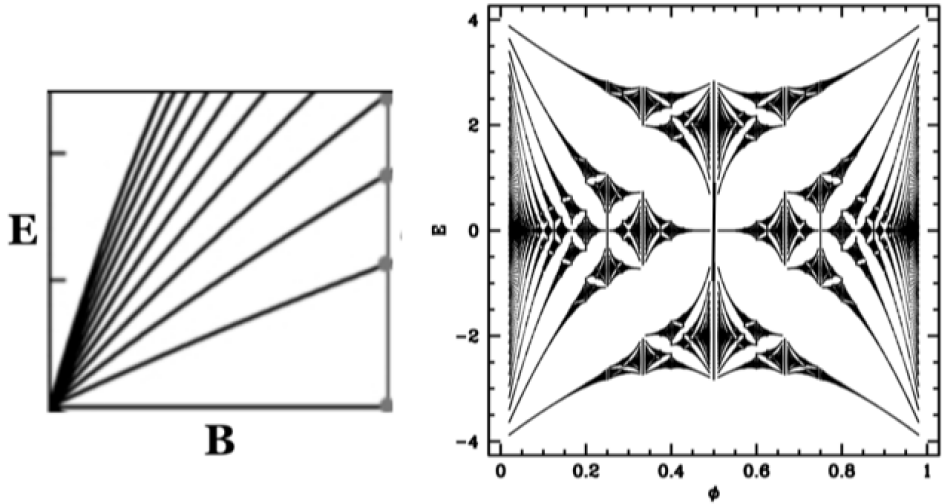} }
\leavevmode \caption{Energy spectra of electrons in two-dimensions in a magnetic field. The left and the right panels respectively show the spectrum for the continuum (vacuum) and  square lattice (crystal) cases.}
\label{B}
\end{figure}
 
The Butterfly Hamiltonian is obtained using a simple model of electrons in a two-dimensional square lattice where electrons can hop only to their neighboring sites.  When subjected to a magnetic field, the Hamiltonian is given by,
 
  \begin{equation*}
 H =  \cos a (k_x -\frac{e}{c}A_x) + \cos a( k_y  - \frac{e}{c} A_y),
 \end{equation*}
 where the magnetic field $\vec{B} = \nabla \times \vec{A}$.  Here $a$ is the lattice constant and $H$ is defined in units of the strength of the nearest-neighbor hopping parameter.
 With the choice of gauge,  $(A_x = 0, A_y = Bx )$, and the wave function $\Psi_{n,m} = e^{i k_y m} \psi_n(k_y)$, the problem effectively  reduces to a one-dimensional system, known as  the Harper's equation\cite{Harper}
 
  \begin{equation}
 \psi_{n+1}+\psi_{n-1} + 2 \cos( 2 \pi n \phi + k_y) \psi_n = E \psi_n.
 \end{equation}
 
 The parameter $\phi = \frac{B a^2}{ \hbar/e}$ is the magnetic flux per unit cell of the lattice, measured in the unit of flux quanta $\hbar/e$.  We note that the Hamiltonian can also be written as\cite{Wil},
 
 \begin{equation}
 H = \cos x + \cos p, \,\,\,\, \ [x, p] = i \phi
 \end{equation}
 
 That is, the butterfly graph lives in a space of energy E and ( effective) Planck's constant $\phi$. (he corresponding continuum Hamiltonian is $ H = p^2+ x^2$.
 
 Figure (\ref{B}) shows the spectrum for both the continuum and the  corresponding lattice model,  highlighing the importance of  role of completing periodicities in a lattice system subjected to a magnetic field. 
 
 \begin{figure}
\resizebox{0.8 \columnwidth}{!}{%
  \includegraphics{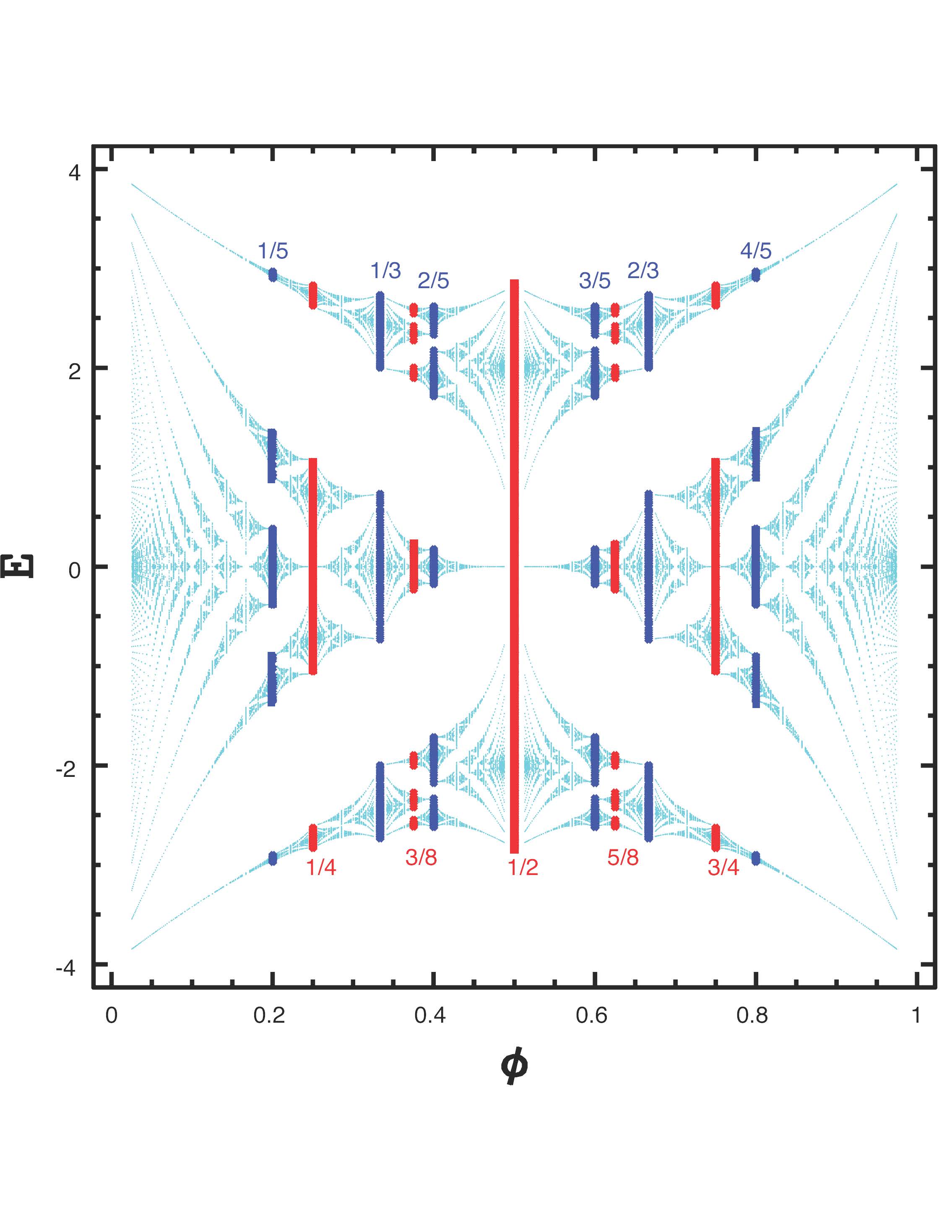} }
\leavevmode \caption{A Hofstadter butterfly graph where we highlight part of the band spectrum for  some rational flux values $p/q$ shown in blue ($q$-odd) and in red ($q$-even).}  
\label{band}
\end{figure}

\subsection{ The Butterfly Spectrum }

For a  flux $\frac{p}{q}$,  the butterfly spectrum consists of  $q$ bands, separated by $(q-1)$ empty spaces -- the gaps that form the wings of the butterfly. Bands are made up of values of energy that are permissible and gaps are values of energy that are forbidden quantum mechanically.
 For $q$ even, the two bands touch  or kiss at the center of the spectrum, that is at $E=0$ as shown in Fig. (\ref{band}).
 For the irrational value of the flux, the spectrum is a Cantor set where the allowed values of  the energy is set of zero measure. This is known as the ``Ten Martini Problem" --
the name was coined by Barry Simon in his 1982 article\cite{Ten}, originating from the fact that Mark Kac has offered ten martinis to anyone who solves it.

Coexisting simplicity and complexity is a norm in  the Hofstadter butterfly.  This is also reflected in  Chamber's formula\cite{Chamber} which states that the energies $E$
 of the Harper's equation, for magnetic flux $\phi= \frac{p}{q}$  depend upon $k_x$ and $k_y$ -- the Bloch vectors in $x$ and $y$ direction, via $\Lambda$  which is equal to the determinant of $H$,
 
 \begin{eqnarray*}
 \Lambda ( k_x, k_y )  & = & 2 \cos ( q k_x) + 2 \cos( q k_y)\\
 &  = & E^q + a_1 E^{q-1} + a_2 E^{q-2}+.......
 \end{eqnarray*}
 Here the coefficients of the polynomial $a_i$ are independent of $k_x$ and $k_y$.
 Analytic expressions below  for the energy dispersions $E(k_x, k_y)$ for a few simple cases illustrate this.

\quad ({\bf a}) \,\, For $\phi = 1$, the energy spectrum consists of a single band given by $E =  2 ( \cos k_x + \cos k_y )$.

\quad ({\bf b}) \,\, For $\phi=1/2$, the two bands having energies $E_+$ and $E_-$ are given by $E_{\pm}=\pm 2 \sqrt{ 1+ \frac{1}{2}( \cos2 k_x + \cos2 k_y)}$.

\quad ({\bf c}) \,\, For $\phi=1/3$, the three bands have energies  $E_{n} =   2\sqrt{2} \cos ( \theta \pm  n \frac{2}{3} \pi)$. Here, $\theta= \frac{1}{3} Arccos[( \cos 3k_x+ \cos 3k_y)/2\sqrt{2}]$, $ n = 0,1,2$.

\quad ({\bf d}) \,\, For $\phi = 1/4$, the energies of four bands are given by the expression $E= \pm \sqrt{4 \pm 2[3+\frac{1}{2} (\cos 4k_x + \cos 4 k_y)]^2}$.

These values of energy are shown in Fig. (\ref{band}) where each energy band is obtained by varying $(k_x,k_y)$ in the Brillouin zone.

\begin{figure}
\resizebox{1.0\columnwidth}{!}{%
  \includegraphics{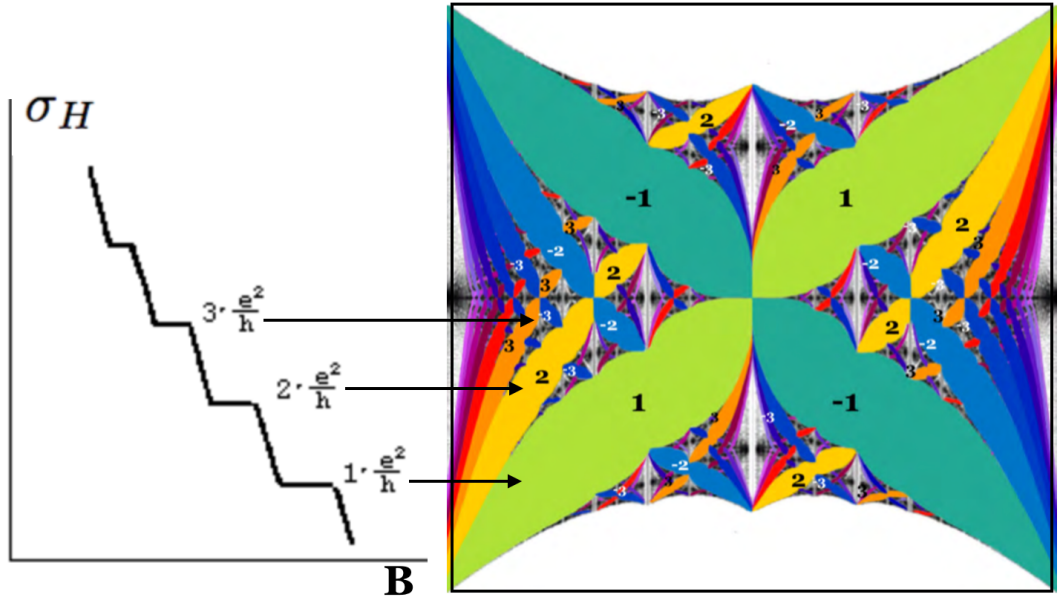} }
\leavevmode \caption{  In this  graph, the gaps of the butterfly are colored. Integer quanta label the gaps, and gaps with the same color represent quantum Hall states with the same quantum number of Hall conductivity.}  
\label{Integer}
\end{figure}

\subsection{Integers of  the Butterfly as  the Quanta of  Hall Conductivity }

Intriguingly,  the parts of the butterfly graph corresponding to the energies of the electrons unacceptable by the rules of quantum mechanics,  become important in the quantum world due to the some  special properties of  
the wave functions.  The integers that label the wings of the butterfly are in fact hidden in the ``geometry" of the wave function.

In $1983$  David Thouless\cite{TKNN} ( for which he was awarded the 2016  Nobel prize in physics prize)  along with his collaborators showed that  the Hall conductivity  $\sigma_H$ can be written as,

\begin{equation}
\sigma_ H=  [\frac{i}{2 \pi} \sum_{n=1}^{n_f}  \int_{T} \{\partial_{k_x}  \psi_n^*  \partial_{ k_y}  \psi_n- \partial_{k_x}  \psi_n  \partial_{ k_y} \psi_n^*\} \, dk_x \, dk_y] \,\ \frac{e^2}{h}.
\label{BC}
\end{equation}

Here $n_f$ represents the number of filled bands as the Fermi energy energy lies in the gaps of the spectrum. The quantity in the square bracket can assume only integer values which can be interpreted as a ``curvature"  known as the Berry curvature and its integrated value is the Berry phase\cite{BP}.  The Hall conductivity $\sigma_H$ represents the Berry phase in units of $ 2 \pi$.  
This rather miraculous result is due to the fact that for any two-dimensional closed manifold, the total Gaussian curvature is always an integer multiple of $2\pi$. This is a generalization of  the Gauss-Bonnet theorem in differential geometry  by Shiing-Shen Chern to quantum systems to describe the geometry of wave functions.

The Berry phase is rooted in the quantum anholonomy where the wave function does not return to its original value after a cyclic loop in some parameter space. We note that  in modern geometry, the local curvature
at a point is defined by the anholonomy of a tangent vector that fails to return to its starting location after undergoing a cyclic journey around an infinitesimal loop at that point. Interestingly, the quantum phenomenon is analogous to the
classical anholonomy that results in net rotation in the  plane of oscillation of a Foucault pendulum after earth rotates through an angle of $2\pi$ radians\cite{BP}.

\begin{figure}
\resizebox{0.75 \columnwidth}{!}{%
  \includegraphics{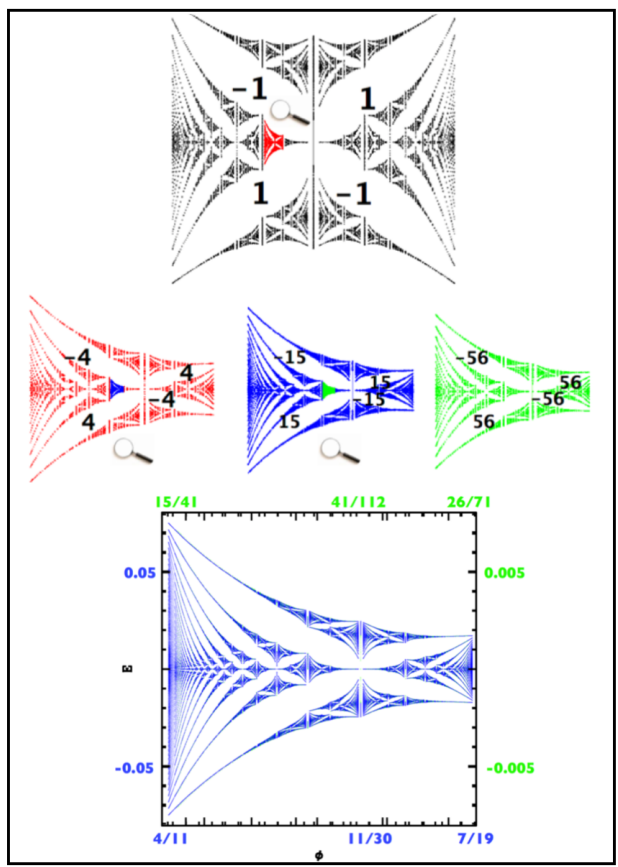} }
  \leavevmode \caption{ Zooming into the butterfly fractal reveals identical patterns at all scales. The red butterfly is a blowup of the red region in the upper black graph. The blue butterfly is a blowup of the blue region in the red graph, and the green butterfly is, in turn, a blowup of the green region in the blue graph.  The lowest panel shows the overlay of the blue and green butterflies implying self-similarity.}\label{Self}
\end{figure}

\begin{figure}
\resizebox{1.0\columnwidth}{!}{%
  \includegraphics{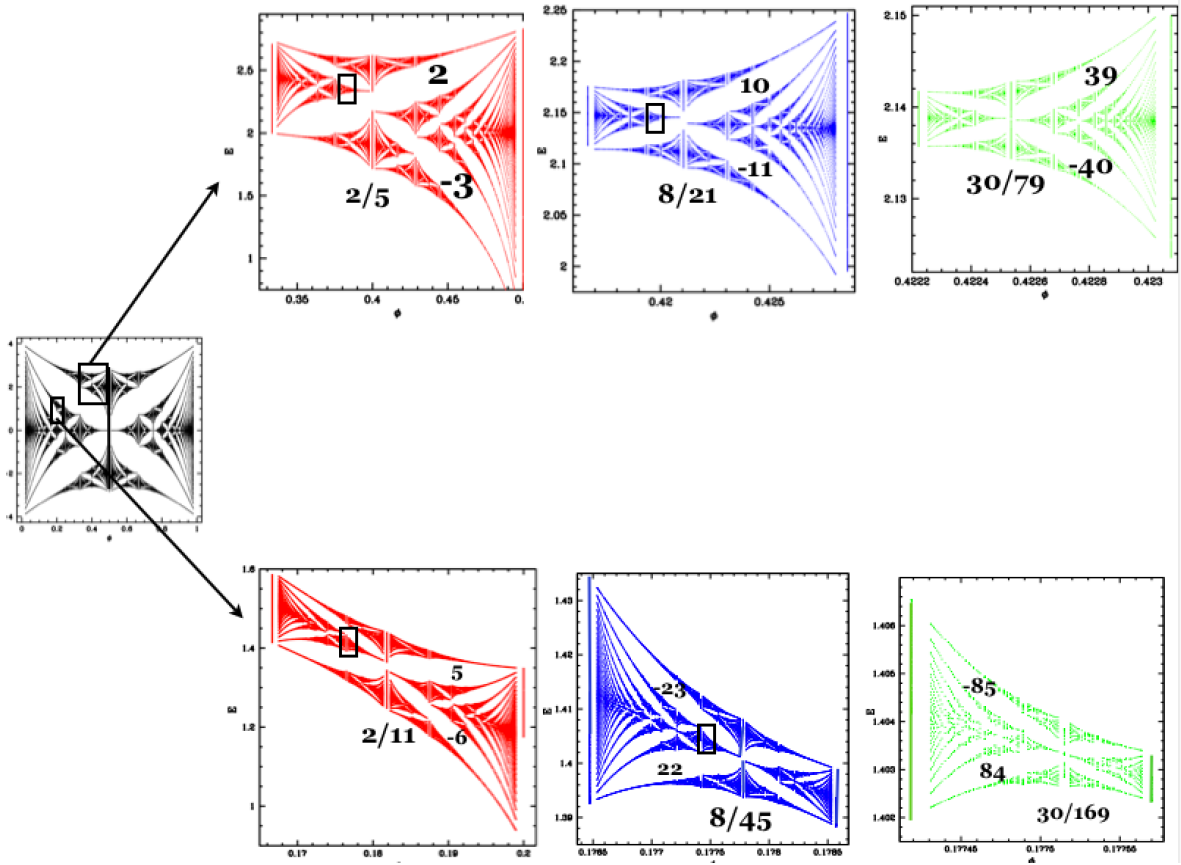}}
\leavevmode \caption{ Two sequence of blowups, each  shown in red, blue and green of the boxed parts of the butterfly on the left (black).  Note that  in both cases we start with a C-cell butterfly  that exhibits no mirror symmetry and the two bands at the center do not kiss. However,
  the images of sequence of blow ups of its central cell evolve into a configuration of horizontal mirror symmetry where the  difference between the two Chern numbers, that is  $\sigma_+ - \sigma_-$  remain constant and the two central bands almost kiss.  Also, the parity of $q_c$ ( even or odd-ness)  is preserved as we zoom into equivalent set of butterflies at smaller and smaller scale.}
\label{Ccell}
\end{figure}

\begin{figure}
\resizebox{1.0\columnwidth}{!}{%
  \includegraphics{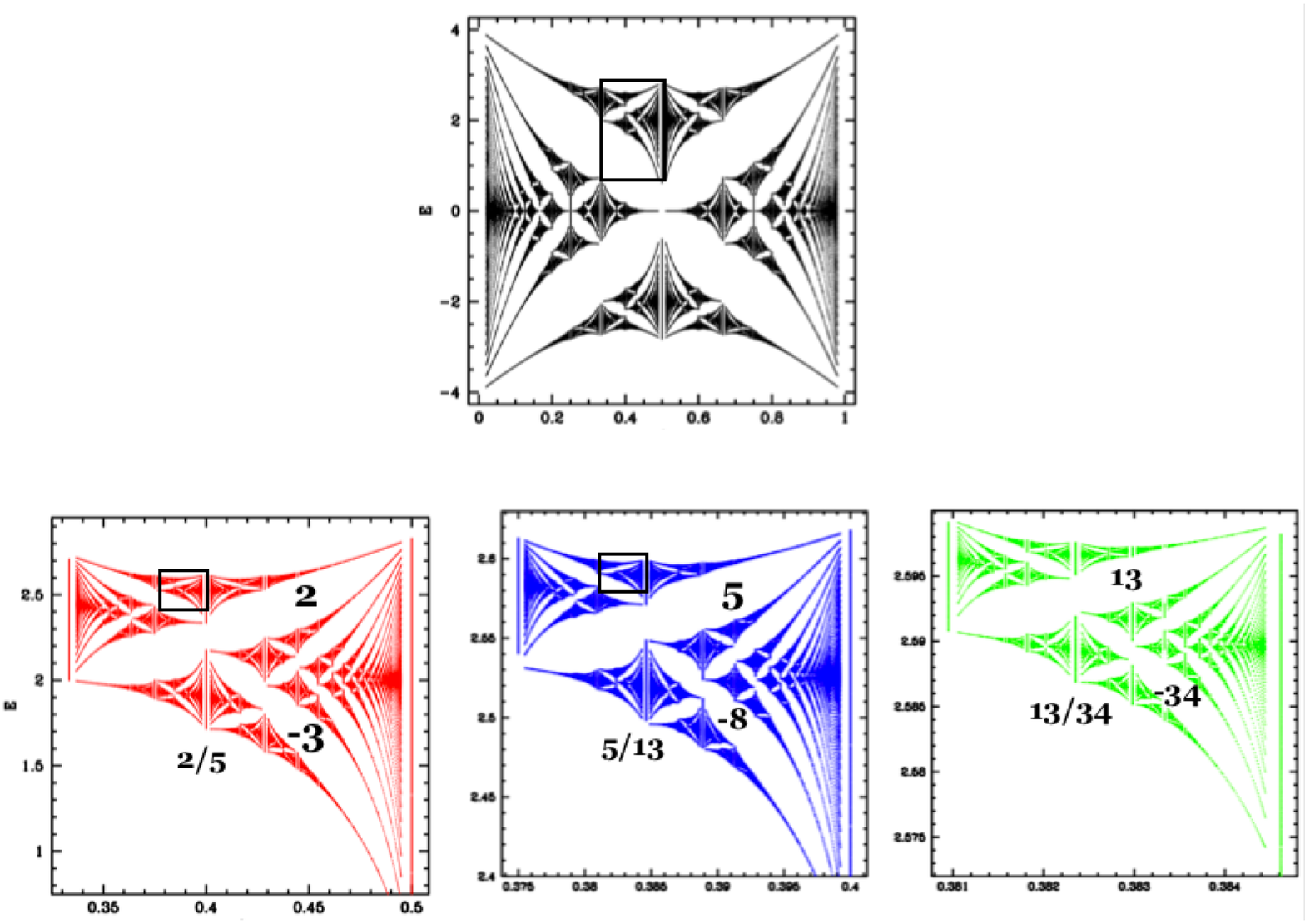}}
\leavevmode \caption{ Analogous to the Fig. (\ref{Ccell}), here is an example of a E-cell  butterfly hierarchy. In sharp contrast to  the blow ups of the C-cell butterflies, these  blow ups  do not evolve into any symmetric configuration, the two central bands do not kiss and and the two Chern numbers drift further and further apart. Also, the parity of $q_c$ is not conserved.}
\label{Ecell}
\end{figure}

\section{ The Butterfly Fractal}

 Figure (\ref{Self})  illustrates  self-similar fractal properties of the energy  spectrum  consisting  of ( distorted ) butterflies at all scales. 
 In every image, all  the four ``wings"  (the gaps)  of a butterfly pass through a common flux value  $\frac{p_c}{q_c}$. This flux value is identified as the center of the butterfly where the spectrum consists of two bands.
 The flux values to the left and the right of this center where the spectrum is  a single band defines the left and the right boundaries of the butterflies.

Figues (\ref{Ccell}) and (\ref{Ecell})  further illustrate the adfinitum images of the whole pattern -- nested set of distorted butterflies, as we zoom into different parts of the spectrum.  We list below some of the key features  of the butterflies as illustrated in these graphs.

\begin{itemize} 

\item Two species of butterflies:\\
The butterfly hierarchies within every butterfly graph can be classified into two categories: the {\it central } and the {\it edge} butterflies which we respectively refer as the C-cell and  the E-cell butterflies.  
Asymptotically, C-cell butterflies evolve into configuration
where the two bands at the butterfly center almost kiss and the butterflies recover horizontal mirror symmetry. This corresponds to   $\sigma_+ \rightarrow \sigma_-$, that is the magnitude of two Chern numbers of its two diagonals  approach each other. 
In sharp contrast to the C-cell butterflies,  the E-cell butterflies evolve
into highly asymmetric image where the two central bands and their Chern number drift further and further apart as seen in Fig. (\ref{Ecell}). 
\item  Butterfly Parity:\\

An important feature of the
 the hierarchical set of  the C-cell butterflies  is the fact that they conserve {\it parity} which we define as  even (odd) when $q_c$ is even (odd).  We note that  butterflies with center at $E=0$ shown in  Fig. (\ref{Self}) are special case of the C-cell butterflies that exhibit  even parity and horizontal mirror symmetry at all scales. These properties are displayed Fig. (\ref{Ccell}).

The E-cell butterfly hierarchies do not conserve parity. That is, as we zoom into the equivalent butterflies, $q_c$ oscillates between an even and  an odd integer.
\end{itemize}

\section{  NUMBER THEORY \& THE BUTTERFLY Graph}

Perhaps, the most important  number theoretical feature of the butterfly graph  is the {\it Farey Relation}. Found empirically, it  links the flux coordinates of the butterfly center and its  boundaries and is given by,

\begin{equation}
\frac{p_c}{q_c}   =  \frac{p_L+p_R}{q_L+q_R}.
\label{bid}
\end{equation}

Here the three rational numbers $\frac{p_L}{q_L}$, $\frac{p_c}{q_c}$ and
$\frac{p_R}{q_R}$  respectively describe the left edge, the center and the right edge of a butterfly on the magnetic flux axis,
Consequently, three fractions  always form a friendly triangle on the Farey tree. That is, 
the butterfly center, its left boundary and the right boundary are all Farey neighbors of each other,
 \begin{equation}
 q_x p_y - q_y p_x = \pm 1,
 \end{equation}
 where $x$ and $y$ refer to any of the two labels  $L$, $c$ or $R$.  As described below, this maps butterflies to Apollonians and the Pythagorean triplets.
 
 Another remarkable feature of the spectrum is that each of the two diagonals of every butterfly is labeled by two  integers $(\sigma, \tau)$ that have number theoretical origin.  They are solutions of 
 a linear Diophantine equation\cite{Dana},
 \begin{equation}
r  =  \sigma p + \tau q. 
\end{equation}

Here $r$ is the gap index,  and for a given rational flux $\phi = \frac{p}{q}$, $ r= 1, 2, ... q-1$.  Given $r$ and $\frac{p}{q}$, there is a unique solution, mod (q),  for $\sigma, \tau$. 
Therefore, every butterfly is characterized by two pairs of integers:   $(\sigma_+, \tau_+)$, and $(\sigma_-, \tau_-)$, labeling its  two diagonals. The $(\sigma_+, \sigma_-)$  are the Chern numbers
-- the topological quantum numbers  $\sigma_H$ in Eq.  (\ref{BC}).  However, the physical significance of $\tau$ remains unknown.

\begin{figure}
\resizebox{1.0\columnwidth}{!}{%
  \includegraphics{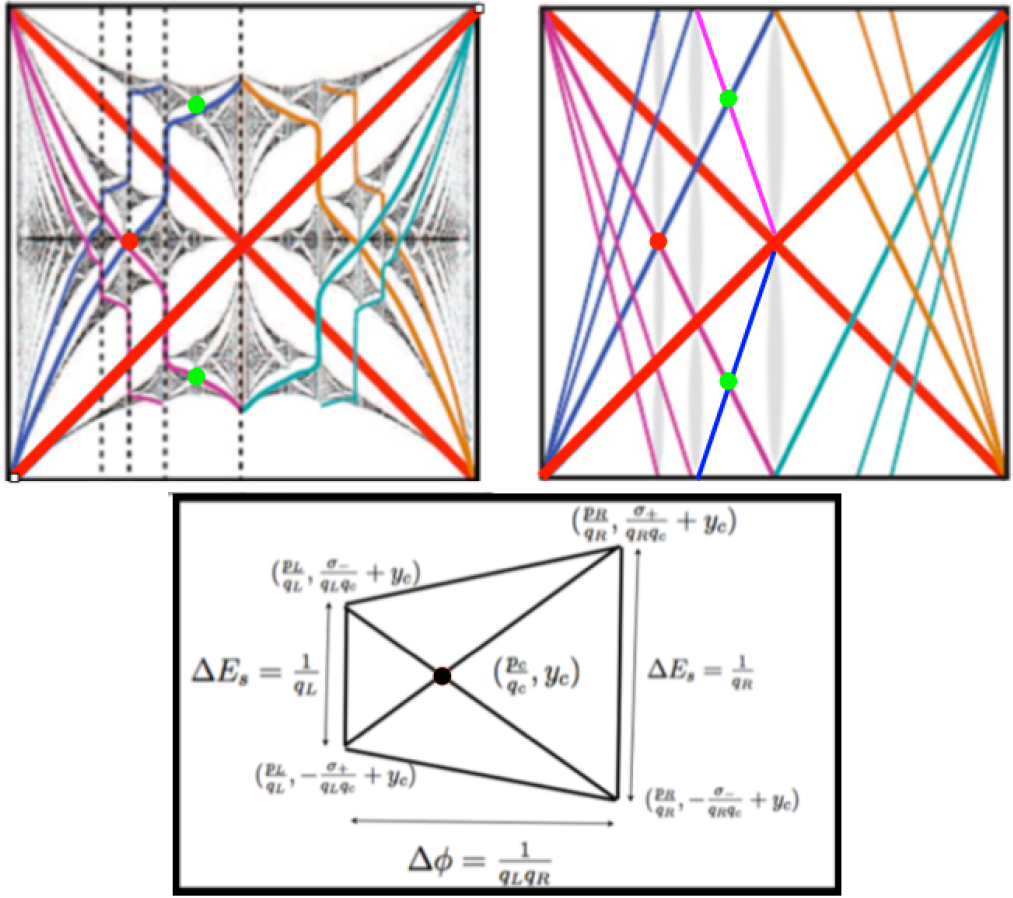} }
\leavevmode \caption{  Upper right graph shows the butterfly skeleton  --- where the energy gaps of the Hofstadter spectrum are simplified down to linear trajectories parametrized as $\rho  \equiv r/q= \sigma \phi + \tau$. In the actual butterfly diagram, the linear trajectories become discontinuous, as is shown by the black vertical lines in between the colored lines as shown by the upper left panel in the graph.  The lower graph shows the trapezoid region in $\rho-\phi$ plane
labeling the coordinates of the butterfly skeleton.
The pair of vertical parallel sides
at flux values $\phi_L$ and $\phi_R$ define the left and the right boundaries of the butterfly. }\label{WD}
\end{figure}

In the $\rho-\phi$ plane that is  equivalent to the $E-\phi$ plane of the spectrum , these four integers 
determine the  butterfly center $(x_c, y_c)$ as \cite{ChernMeet}.  
\begin{eqnarray}
x_c   & =  &  \frac{ \tau_+ + \tau_-}{ \sigma_+ + \sigma_-}  \equiv \frac{p_c}{q_c}\\
 y_c & =  & \frac{1}{2}[(\frac{ \sigma_+ - \sigma_-}{ \sigma_+ + \sigma_-} )(\tau_+ + \tau_-) -  (\tau_+ - \tau_-)]
\label{Bcoordt}
\end{eqnarray}

\begin{figure}
\resizebox{0.7 \columnwidth}{!}{%
  \includegraphics{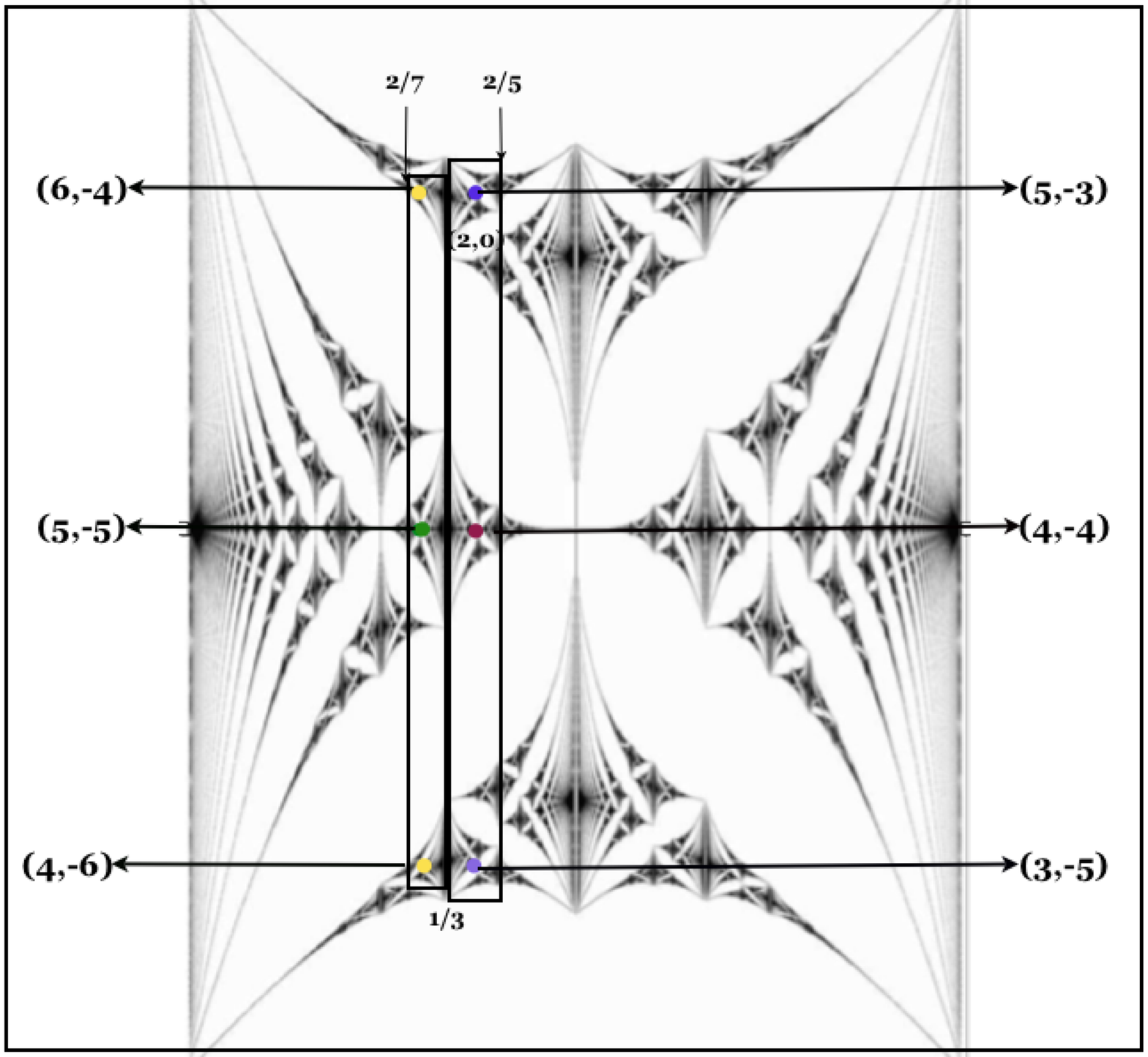}}
\leavevmode \caption{  Figure shows two  cases where  three distinct butterflies  share the same  flux interval.  Red and blue dot butterflies share the interval$1/3 \le \phi \le 2/5$. The  green  and yellow dot butterflies share the interval
$2/7 \le \phi \le 1/3$.  The pair of integers  marked with arrows attached to the butterfly  are the two Chern numbers associated with the two diagonals of the butterfly
show that two distict butterflies that share flux interval have different Chern numbers.}
\label{onemany}
\end{figure}

We note that the  $\{ \sigma_+, -\sigma_-, - \tau_+ , \tau_-\}$  completely specifies a  butterfly. This is because, given the  $x$-coordinate of the butterfly center $x_c = \frac{p_c}{q_c}$, its flux boundaries are  the  Farey neighbors of $ \frac{p_c}{q_c}$.  As shown in Fig. (\ref{WD}),  the four lines -- the two diagonals  $y= \sigma_+ x - \tau_+$ and  $y= -\sigma_- x + \tau_-$ and  the two vertical lines at  the Farey neighbors of $\frac{p_c}{q_c}$ intersect at four points -- defining a trapezoidal region that we identify as the skeleton butterfly.

Fig. (\ref{onemany}) shows `` sibling butterflies" -- that is group of butterflies  that share a magnetic flux interval.  Each member of this family of butterflies  have different Chern numbers.
Figure emphasizes that  the magnetic flux at the center along with the the two Chern numbers of the diagonal wings uniquely characterize a butterfly.

For  characterizing the C-cell butterflies, just two integers such as $(q_L, q_R)$ or $p_c$ and $q_c$ are needed. The  topological quantum numbers $(\sigma, \tau)$ can be determined in terms of  these integers. For example, for butterflies with centers at $E=0$, $\sigma_+ = \sigma_- = \frac{q_c}{2}$ and $\tau_{\pm}  =  \frac{p_c \mp  1}{2}$.

\begin{figure}
\resizebox{0.75\columnwidth}{!}{%
  \includegraphics{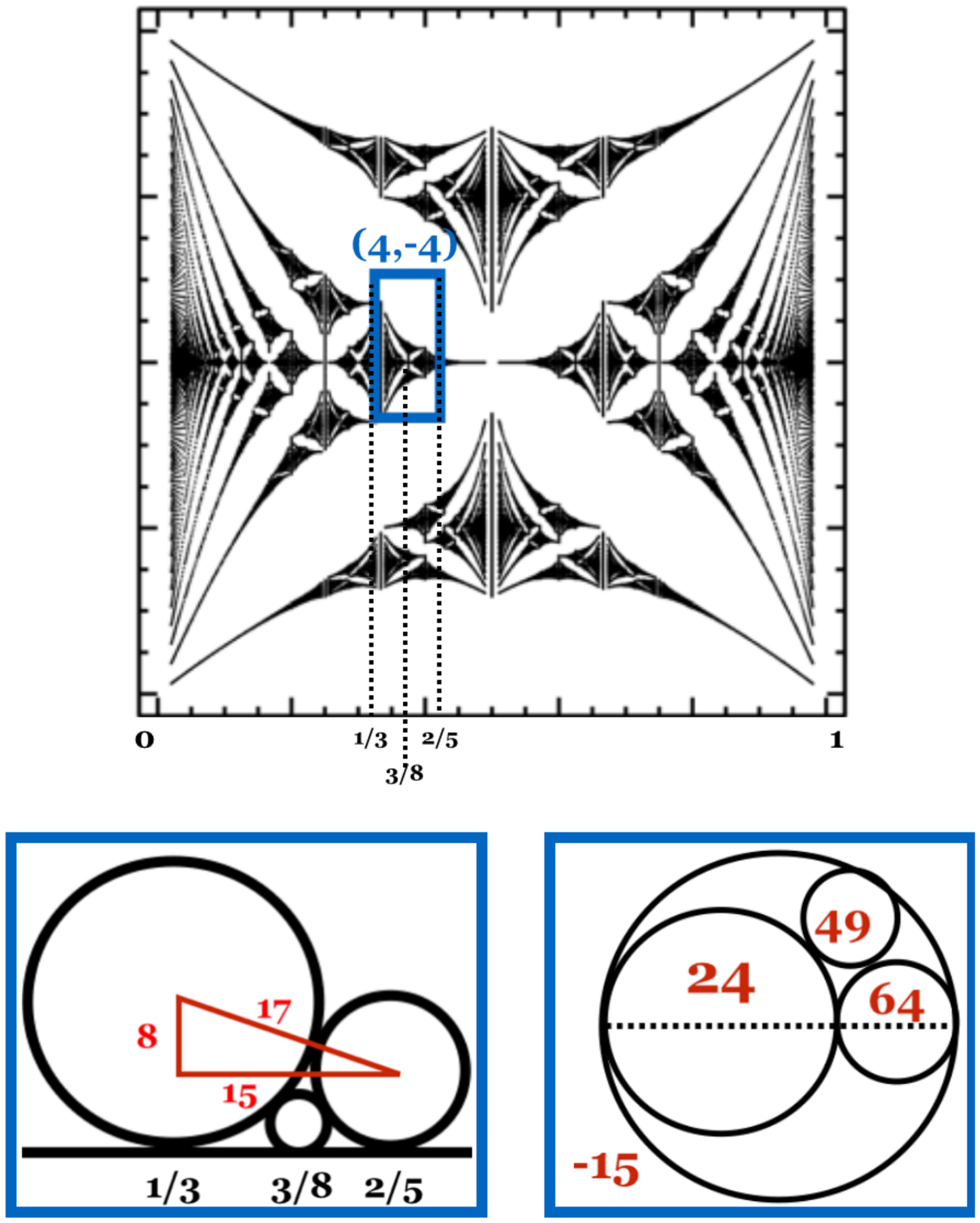} }
\leavevmode \caption{ Figure illustrates how a symmetrical butterfly centered at $\phi=3/8$ is  represented  by a $\cal{PT}$  where the denominators $(5,3)$ of the flux values at the boundary provide the Euclid parameters.  Figure (lower right)  shows the corresponding  Apollonian that  represents the butterfly.}
\label{evenodd}
\end{figure}

\section{ Pythagoras -Apollonian - Butterfly Meet  }

We now consider only the C-cell butterflies. We will discuss both the even parity cases where the butterflies have their center at $E=0$ and the odd parity butterflies whose centers reside off the $E=0$ line. 
\begin{figure}
\resizebox{0.6\columnwidth}{!}{%
  \includegraphics{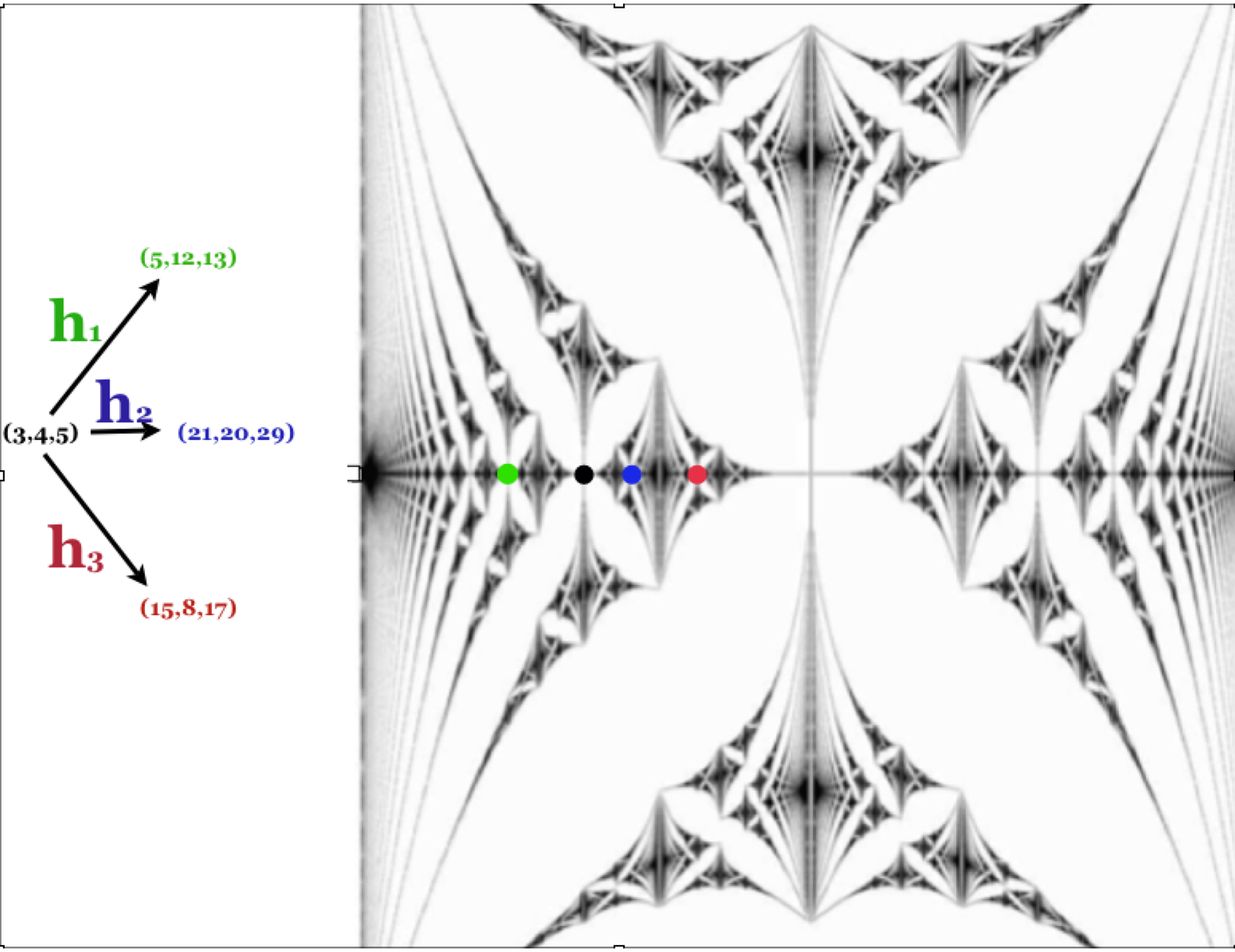} }
\leavevmode \caption{ For C-cell, this figure illustrates the relation between the butterfly nesting and the Pythagorean tree: Starting with a butterfly centered at flux $1/4$ ( shown with a red dot ) which is represented by the Pythagorean triplet
$(3,4,5)$,  figure illustrates the butterflies ( shown with red dotes )  generated by the three matrices $h_1, h_2, h_3$.}
\label{bpt}
\end{figure}

A direct consequence of the Farey relation ( see Eq. (\ref{bid}) )  is the fact that representation of  butterfly flux coordinates $(\frac{p_L}{q_L}, \frac{p_c}{q_c}, \frac{p_R}{q_R})$ as Ford circles  associates the C-cell butterfly with a configuration of three kissing Ford circles. This  configuration in turn maps to a Pythagorean triplet as illustrated in Figures  (\ref{ap}) and (\ref{apt}). The Euclid parameter for mapping butterfly to the Pythagorean triplet are $(q_L, q_R)$.

For butterflies with their centers at $E=0$ ,  $q_c$ is always even.  For such {\it even- parity} butterflies,  the corresponding Pythagorean triplet  are:
 
  \begin{eqnarray}
  (n_x,n_y,n_t) & = &  ( q_L q_R, \frac{1}{2}(q_R^2-q_L^2),  \frac{1}{2}(q_R^2+q_L^2))
  \label{bspinor}
  \end{eqnarray} 

For  C-cell butterflies whose center do not reside at $E=0$,  $q_c$ is odd. Such  {\it odd-parity} butterflies can be described by a ``dual" Pythagorean tree --
a tree where the legs  of the right triangle ( that is $n_x$ and $n_y$ ) are interchanged. The Pythagorean triplets for these butterflies where $q_R$ and $q_L$ have opposite parity are obtained by multiplying by a factor of $2$ in Eq. (\ref{bspinor}). Fig. (\ref{evenodd}) shows an example of the mapping of a C-cell butterfly to a configuration of three kissing Ford circles and to a Pythagorean triplet.

 We note that the  integers $n_x$ and $\kappa_0$ determine  the horizontal size $\Delta \phi$ of a butterfly:

\begin{equation}
\Delta \phi = |\frac{p_R}{q_R}-\frac{p_L}{q_L}| = \frac{1}{q_L q_R} =   \frac{1}{n_x}=\frac{1}{\kappa_0}
\label{bsize}
\end{equation}

\section{ Butterfly Recursions}

An important consequence of the C-cell butterfly-$\cal{PT}$ mapping is that  the Pythagorean tree  shown in Fig. (\ref{PTnew}) describes the recursive pattern underlying such butterflies. 

Figure  (\ref{bpt}) shows how Lorentz transformations $h_1, h_2, h_3$  generate  the butterflies with center at $E=0$.  Each  butterfly hierarchy is associated with a unique path in the tree, characterized by a string of the $h$ matrices.
For C-cell butterflies whose centers do not reside at $E=0$, and have odd-parity are described by the dual Pythagorean  tree with $n_x, n_y$ interchanged. In other words,  Pythagorean tree and its dual fully encode the recursions of all the butterflies that conserve parity. 

Below we give examples of butterfly recursions -- the renormalization equations that connect two consecutive butterflies of an hierarchy and their fixed points lead to self-similar patterns. The examples are grouped into three distinct classes  labeled as $C(\frac{1}{3} \leftrightarrow \frac{1}{2})$,
$C(\frac{1}{3} \leftrightarrow \frac{1}{4})$ and $E(\frac{1}{2} \leftrightarrow 0)$. The first two classes  respectively describe  C-cell butterflies that are confined to the flux intervals $( \frac{1}{3} \le \phi \le  \frac{1}{2})$,
$(\frac{1}{3} \le \phi \le  \frac{1}{4})$ and the last class describes E-cell butterflies in the flux interval $(\frac{1}{2} \le \phi \le  0) $. Each group contains infinity of hierarchies, labeled with an integer ``k".

\begin{itemize}
\item (I) $C(\frac{1}{3} \leftrightarrow \frac{1}{2})$\\
 
These butterfly hierarchies correspond to the paths $ \hat{T}_k =  h^k_3 h_1 $ ( $k=1,2,3...$ )  in the Pythagorean tree, where  each $k$ value corresponds to a distinct hierarchy characterized by a universal scaling.
From Eq. (\ref{h2}), we obtain,

 \begin{equation}
\left( \begin{array}{c}  q_R(l+1) \\   q_L(l+1) \\ \end{array}\right) = \left( \begin{array}{cc} 1+k   & 2+k  \\  k  &1+ k   \\  \end{array}\right)
 \left( \begin{array}{c}  q_R(l) \\   q_L(l) \\ \end{array}\right) 
 \label{s2}
  \end{equation}

\item (II)  $C(\frac{1}{3} \leftrightarrow \frac{1}{4})$\\
 
These hierarchies correspond to the paths $ \hat{T}_k =  h^{k-1}_3 h_2 h_2$ in the Pythagorean tree.

 The recursion  are given by, 
 
 \begin{equation}
\left( \begin{array}{c}  q_R(l+1) \\   q_L(l+1) \\ \end{array}\right) = \left( \begin{array}{cc} 3k+2   &  k+1  \\  3k-1 & k   \\  \end{array}\right)
 \left( \begin{array}{c}  q_R(l) \\   q_L(l) \\ \end{array}\right) 
 \label{s2}
  \end{equation}
 
\item (III) $E(\frac{1}{2} \leftrightarrow 0)$\\

This example considers recursions for E-cell butterflies. Empirically found recursions for this class  of butterflies are given by the following equation,

\begin{equation}
\left( \begin{array}{c}  q_R(l+1) \\   q_L(l+1) \\ \end{array}\right) = \left( \begin{array}{cc} 1   &  k  \\  1 & k+1   \\  \end{array}\right)
 \left( \begin{array}{c}  q_R(l) \\   q_L (l) \\ \end{array}\right) 
 \label{ss}
  \end{equation}
  
  \end{itemize}
  
  Table (2) summarizes the scaling  $\zeta$  for the three classes described above. Emergence of orderly number theoretical pattern underlying the butterfly scalings 
  is quite intriguing.  Appendix provides some additional details about the renormalization flows for the C-cell butterflies.
  
  The E-cell butterflies  are not described by the Pythagorean tree. It is  a conjectured that their recursive pattern is
 encoded in  a tree of Pythagorean quadruplets-- a  subject that is  currently under investigation.

   \begin{table}
\begin{tabular}{| c | c | }
\hline
$\hat{T}$ \,\, &   $\zeta$ \,\    \\ \hline
$C(\frac{1}{3} \leftrightarrow \frac{1}{2})$  \,\, & $[2k+1; \overline{1,2k}] $\\
\\\hline
$C(\frac{1}{3} \leftrightarrow \frac{1}{4})$  \,\, &  $[4k+1; \overline{1,4k}] $\\ 
\\
\hline
$E(\frac{1}{2} \leftrightarrow 0)$  \,\, &   $ [(k+1), \overline{1, k}]  $ \\ 
\\
\hline
\end{tabular}
\caption{  The table summarizes the scaling factor $\zeta$ -- the asymptotic scaling for the topological integers and the magnetic flux intervals for the three classes of hierarchies described above. }
\end{table}

 \section{  Mathematical Kaleidoscope }
 
 It turns out that the hierarchy $h_3 h_1$  described above  also describes the hierarchy of the integral Apollonian with three-fold symmetry. That  is  $\kappa_1=\kappa_2=\kappa_3=\kappa$,  corresponding to  the Apollonian that has perfect three-fold symmetry. The ratio of the curvatures of the innermost and outermost circles ( denoted respectively
as $\kappa^s_+, \kappa^s_-$ is given by ,
 $\frac{\kappa_+}{\kappa_-} \,\, =\,\, \frac{2+\sqrt{3}}{2-\sqrt{3}} \,\, = \, \,(2+\sqrt{3})^2$. ( See Eqs. \ref{DT}, \ref{DT1})
The fact that the ratio of these two curvatures is irrational shows that there is no {\it integral} Apollonian gasket possessing exact three-fold symmetry. However, it is asymptotically approached as one descends deeper and deeper into the  hierarchies of the Apollonian. It is fascinating that
such Apollonians  are l kaleidoscopes in which the image of the first four circles is reflected again and again through an infinite collection of curved mirrors.
 In particular, $\kappa_+$ and $\kappa_-$ are ``mirror images" through a circular ``mirror" that passes though the tangency points of $\kappa^s_1$, $\kappa^s_2$ and $\kappa^s_3$.

\begin{figure}
\resizebox{1.0\columnwidth}{!}{%
  \includegraphics{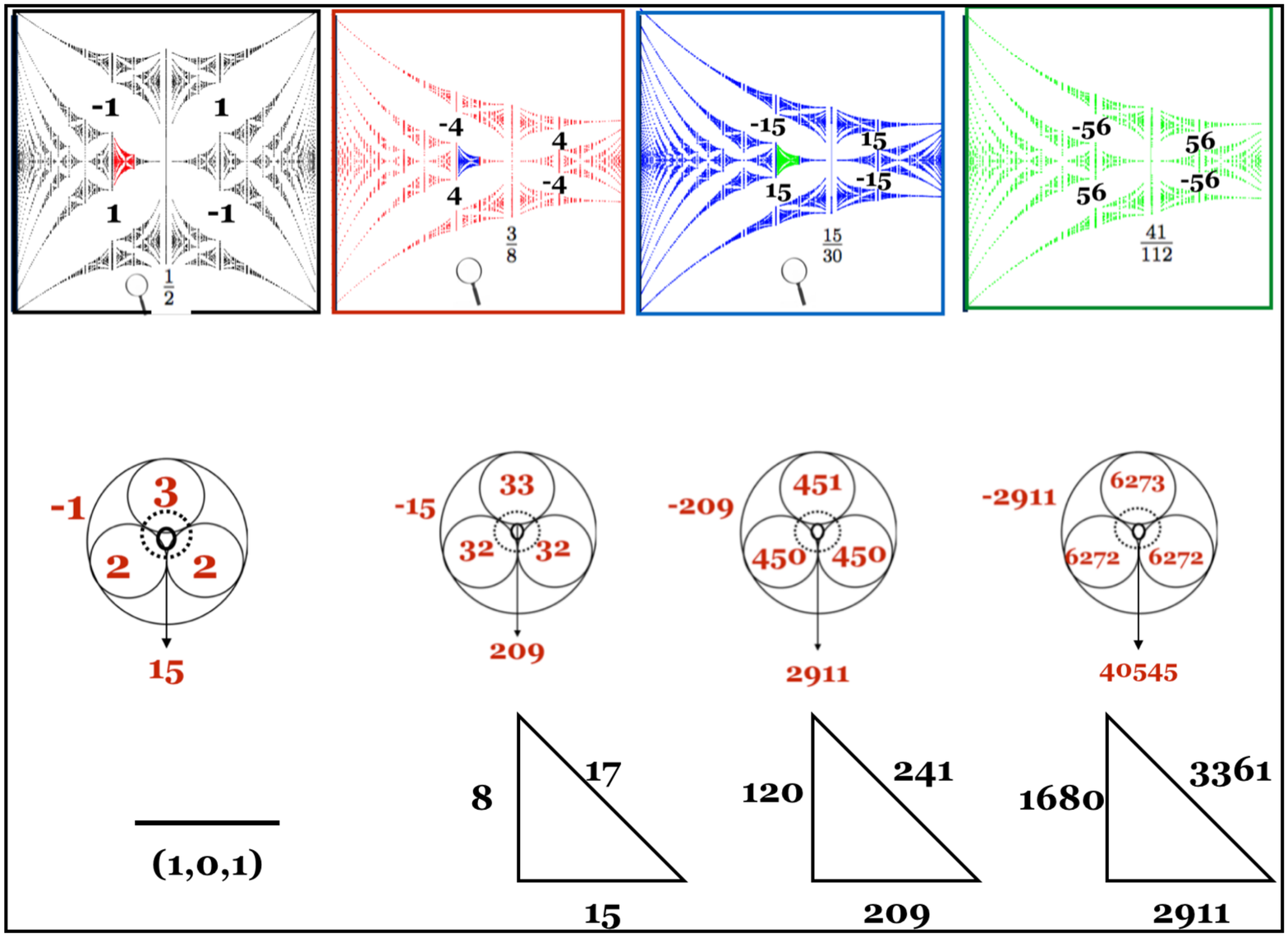}\\
    \includegraphics{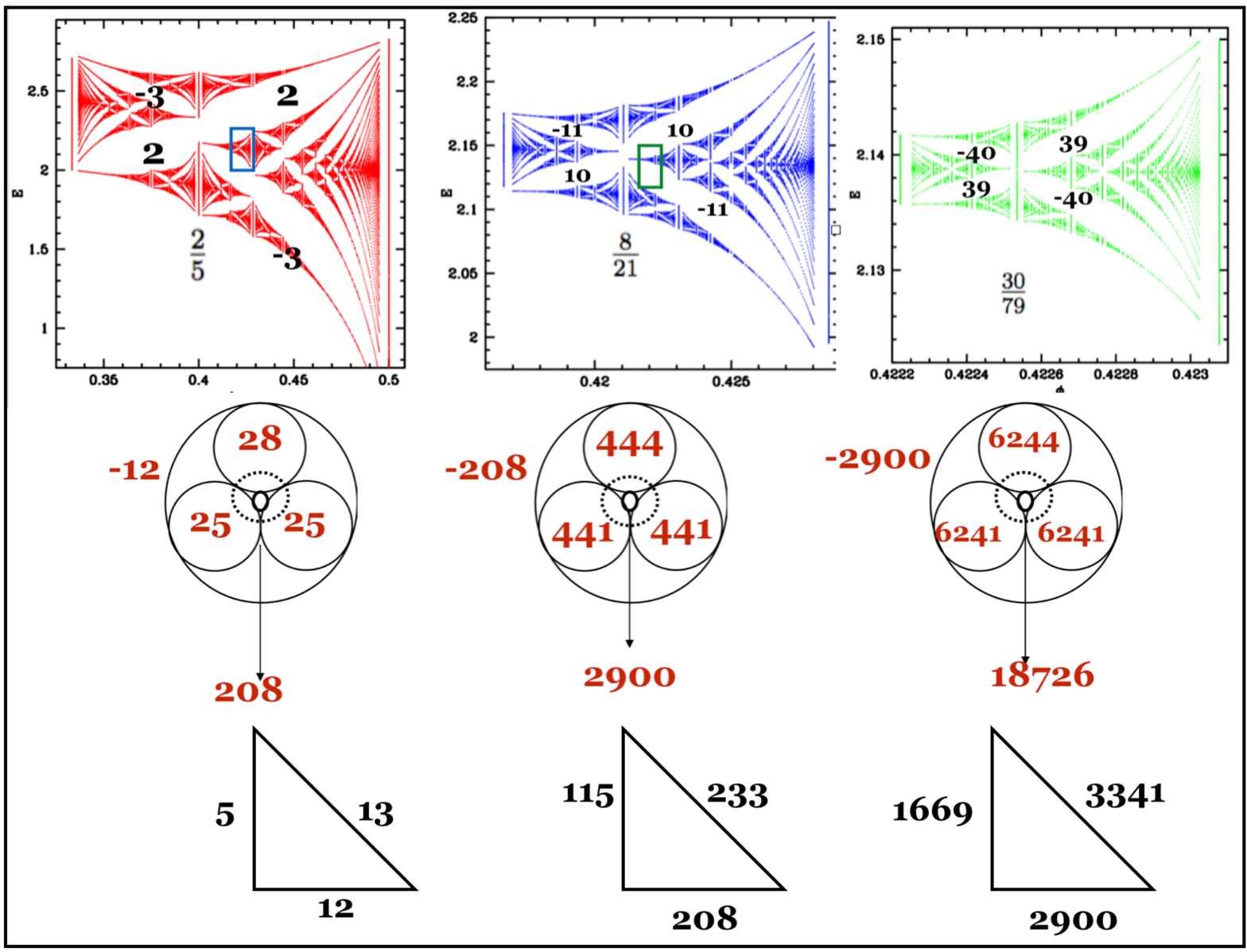} }
\leavevmode \caption{ Two examples of  $\hat{T}= h_3 h_1$ hierarchies  ( starting with two different butterflies )  that exhibit asymptotic three-fold symmetry where the blowups are the kaleidoscopic images. These are characterized by $|\kappa_2-\kappa_3| = |2n_y-n_t|$ remaining invariant at every step of the recursions. This invariant is equal to $1$ and $3$ for the  two hierarchies shown in the left and the right  panels.}
\label{hss}
\end{figure}

 The key to this three-fold symmetry  is the invariants  $| \kappa_2^s-\kappa_3^s|$ of the Apollonians that becomes equal to the invariants $d$ that appears in the Pell's equation. 
This accidental coincidence as displayed in Fig (\ref{hss})  is  a special feature of the $h_3 h_1$ hierarchy making these C-cell butterflies a mathematical kaleidoscope. The emergence of a kaleidoscope
 and the underlying hidden three-fold symmetry in the spectrum of a square lattice is another intriguing number theoretical aspect of the spectrum.
\section{Conclusion}
The butterfly fractal, a 
marvelous example of a physical incarnation of apparently abstract mathematics, is highly complex beast. Recognizing  simple and  familiar mathematical  entities hidden in the  graph is a fascinating aspect of  the butterfly story
that is narrated here.  
Intriguingly, nature chooses neighboring fractions in the Farey tree as the`` building blocks"  to  design  the Hofstadter landscape.  Supplemented by topological integers, these friendly fractions create two species of butterflies --  the  symmetric  and the  asymmetric butterflies. The nesting properties of the symmetric species are encoded in the Pythagorean tree.
Number theoretical  ``ornament"  that embellishes the butterfly is also seen in asymmetric butterfly hierarchies . The emergence of quadratic irrational with continued faction entries consisting of $1$  and $n$ is  pointing towards some deep hidden characteristic of  the butterfly graph that remains unknown.
 ``The unreasonable effectiveness of mathematics in the physical sciences",
(quote from Eugene Wigner), is  a testimony to the inherent simplicity and orderliness that pervades the  fundamental science.

The number theoretical analysis of the butterfly graph as presented here perhaps represents only the tip of the iceberg as 
 recursions  describing asymmetric butterflies hierarchies have  been barely touched. A formulation where every butterfly in the graph can be mapped to an Apollonian and a Pythagorean quadruplet and the recursive structure of the entire graph is described by a tree of Pythagorean quadruplets is currently being investigated. Finally, we note that  our analysis has not touched the energy scaling of the butterfly graph as it falls outside the pure number theoretical domain\cite{Wil}, although number theory dictates this important characterization of the graph.

The butterfly where order and complexity coexist,  remains in many ways a profound enigma. What is particularly fascinating about the butterfly is how both fractality, which is rooted in two competing periodicities, and topology  which is
quintessentially quantum in nature, are interwoven in it. Experimentalists who have seen glimpses of the butterfly in various laboratories\cite{Expt} believe that the study of the butterfly offers the possibility of discovering materials with novel exotic properties that are beyond our present imagination. Who knows how many more mysteries and hidden treasures are yet to be discovered in the butterfly graph.

\subsection{Acknowledgments}

It is a great pleasure to thank Richard Friedberg for some new insights into the number theory  and for pointing out the relationships between the Pythagorean tree and the Pell's equation.
Useful discussions with Jerzy Kocik are gratefully acknowledged.

\section{ Appendix}

Using  $C(\frac{1}{3} \leftrightarrow \frac{1}{2})$ as an example, below we provide further details about the renormalization equations and the scaling properties.

Firstly, we note that the  eigenvalues  of the matrix  $\hat{T}$ are,

\begin{equation}
E_k =  1+k \pm  \sqrt{(1+k)^2-1}
\label{e}
\end{equation}

As described below, these eigenvalues determine the scaling $\zeta$ for the butterfly.

The recursion relations for $(q_L,q_R)$ are given by,

\begin{equation}
\left( \begin{array}{c}  q_R(l+1) \\   q_L(l+1) \\ \end{array}\right) = \left( \begin{array}{cc} 1+k   &  2+k  \\  k & 1+k   \\  \end{array}\right)
 \left( \begin{array}{c}  q_R(l) \\   q_L (l) \\ \end{array}\right) 
 \label{h31}
 \end{equation}
 
 As $l \rightarrow \infty$, we obtain the following,
 
 \begin{eqnarray}
 \frac{q_x(l+1)}{q_x(l) } & \rightarrow & 1+k + \sqrt{(1+k)^2-1},\,\,\  x=0, 1, c \\
 \frac{q_R(l)}{q_L(l)} & \rightarrow & \sqrt{\frac{2+k}{k}}\\
\frac{ \Delta \phi(l)}{\Delta \phi(l+1)} & \rightarrow & \left [1+k + \sqrt{(1+k)^2-1}\right ]^2,
\label{phi}
 \end{eqnarray}
 
 In principle, it is possible to write the corresponding $\phi$ recursions. For example, for $k=1$, that is for $h_3 h_1$ we have $q_L = p_c$. Therefore, we can write,
 
 \begin{equation}
\left( \begin{array}{c}  q_c(l+1) \\   p_c(l+1) \\ \end{array}\right) = \left( \begin{array}{cc} 3   &  2  \\  1 & 1   \\  \end{array}\right)
 \left( \begin{array}{c}  q_c(l) \\   p_c (l) \\ \end{array}\right) 
 \label{hc}
 \end{equation}

Therefore, the Eq. (\ref{h31}) can be transformed into a recursive equation for $\phi_c = \frac{p_c}{q_c}$,

\begin{eqnarray}
\frac{q_R(l+1)}{q_L(l+1)} & =  & \frac{2q_R(l)+3q_L(l)}{q_R(l)+2q_L(l)}\\
\frac{q_c(l+1)-p_c(l+1)}{p_c(l+1)} & = & \frac{q_c(l)-p_c(l)}{p_c(l)} \\
\phi_c(l+1) & = & \frac{1}{2+\frac{1}{1+\phi_c(l)}}
\end{eqnarray}

For butterflies with center at $E=0$, We can now write down the recursions for the topological integers, $(\sigma, \tau)$,\\

 \begin{equation}
\left( \begin{array}{c}  \sigma(l+1) \\   \tau(l+1) \\ \end{array}\right) = \left( \begin{array}{cc} 3   &  1  \\  2 & 1   \\  \end{array}\right)
 \left( \begin{array}{c}  \sigma(l) \\   \tau (l) \\ \end{array}\right) 
 \label{hc}
 \end{equation}
 
 where $\sigma= \sigma_+ = \sigma_-$ and $\tau = \tau_+ + \tau_-$, $\tau_+ = (\tau-1)/2, \tau_- =  (\tau+1)/2$\\ 

 \begin{eqnarray}
   \frac{q_R(l+1)}{q_R(l)}    \rightarrow    \frac{q_L(l+1)}{q_L(l)}  \rightarrow  2+ \sqrt{3} = [3; \overline{1,2}] 
  \end{eqnarray}
  
 This is the scalings associated with topological quantum numbers  $\sigma_+=\sigma_-=\frac{q_R+q_L}{2}$  and  its square describes  scaling of the magnetic flux interval $\Delta \phi$ are ( See Eq. (\ref{bsize}) ),
  
    \begin{equation}
     \frac{ \sigma(l) }{\sigma(l+1)} \rightarrow (2+ \sqrt{3}),\
 \frac{\Delta \phi (l)}{\Delta \phi(l+1)}  \rightarrow (2+ \sqrt{3})^2
\end{equation}

\end{document}